\documentclass[12pt]{article}
\usepackage{amsmath}
\usepackage{graphicx}
\usepackage{enumerate}
\usepackage{natbib} 
\usepackage{url} 
\usepackage{pdflscape}
\usepackage{setspace}
\usepackage{multirow}
\usepackage{hyperref}
\usepackage{amsthm}
\usepackage{amsfonts}
\usepackage{bbm}
\RequirePackage[dvipsnames]{xcolor}
\usepackage{mathptmx}
\newcommand{\blind}{0}

\numberwithin{equation}{section}

\newcommand{\RR}{\mathbb{R}}

\newcommand{\RC}[2]{\mathrm{RC}_{#2}(#1)}

\newtheorem{property}{Property}[section]

\theoremstyle{remark}

\theoremstyle{definition}

\begin{document}

\def\spacingset#1{\renewcommand{\baselinestretch}%
{#1}\small\normalsize} \spacingset{1}

\if0\blind
{
  \title{\bf New estimation methods for extremal bivariate return curves}
  \author{C. J. R. Murphy-Barltrop$^{1*}$, J. L. Wadsworth$^2$ and E. F. Eastoe$^2$\\
\small $^1$STOR-i Centre for Doctoral Training, Lancaster University LA1 4YR, United Kingdom \\
\small $^2$Department of Mathematics and Statistics, Lancaster University LA1 4YF, United Kingdom \\
\small $^*$Correspondence to: c.barltrop@lancaster.ac.uk}
\date{\today}
  \maketitle
} \fi

\if1\blind
{
  \bigskip
  \bigskip
  \bigskip
  \begin{center}
    {\LARGE\bf New estimation methods for extremal bivariate return curves}
\end{center}
  \medskip
} \fi

\bigskip
\begin{abstract}
    In the multivariate setting, estimates of extremal risk measures are important in many contexts, such as environmental planning and structural engineering. In this paper, we propose new estimation methods for extremal bivariate return curves, a risk measure that is the natural bivariate extension to a return level. Unlike several existing techniques, our estimates are based on bivariate extreme value models that can capture both key forms of extremal dependence. We devise tools for validating return curve estimates, as well as representing their uncertainty, and compare a selection of curve estimation techniques through simulation studies. We apply the methodology to two metocean data sets, with diagnostics indicating generally good performance. 
\end{abstract}
\noindent%
\vfill

\newpage
\spacingset{1.8}

\section{Introduction} \label{Sec1}
\subsection{Univariate extremal risk measures}
Statistical analysis of extreme values is important in a wide range of environmental contexts, from the modelling of wildfires to flood risk assessment. The two most common approaches to modelling the extreme behaviour (or tail) of a single variable are block maxima and peaks over threshold \citep{Coles2001}. For the latter, which is more popular in practice, the generalised Pareto distribution (GPD) is used to model exceedances of some high threshold. This is justified through the Pickands-Balkema-de Haan theorem \citep{Balkema1974,Pickands1975}; given a random variable $X \sim F_X$ satisfying certain conditions, there exists a normalising function $c(u)$ such that
\begin{equation} \label{eqn:gpd_cdf}
    \Pr \left(\frac{X - u}{c(u)} \leq x \;  \Big\vert \; X>u\right) \to G(x) := 1 - \left\{1 + \frac{\xi x}{\sigma} \right\}_+^{-1/\xi}, \hspace{.5em} x > 0, \; (\sigma,\xi) \in \RR^+ \times \RR,
\end{equation}
as $u \to x^F := \sup\{x : F_X(x) < 1 \}$. Here, $G$ is the cumulative distribution function of a GPD, with scale and shape parameters, $\sigma$ and $\xi$, respectively, and $z_+ = \max(0,z)$. The shape parameter $\xi$ determines the behaviour of the tail: the cases $\xi<0$, $\xi = 0$ and $\xi>0$ correspond to bounded, exponential and heavy tails, respectively. Given a sufficiently high threshold $u$, we assume the statistical model $X - u \mid X>u \sim \text{GPD}(\sigma,\xi)$. A detailed discussion of peaks over threshold modelling can be found in \citet{Coles2001}. 

\begin{figure}[ht]
    \centering
    \includegraphics[width=\textwidth]{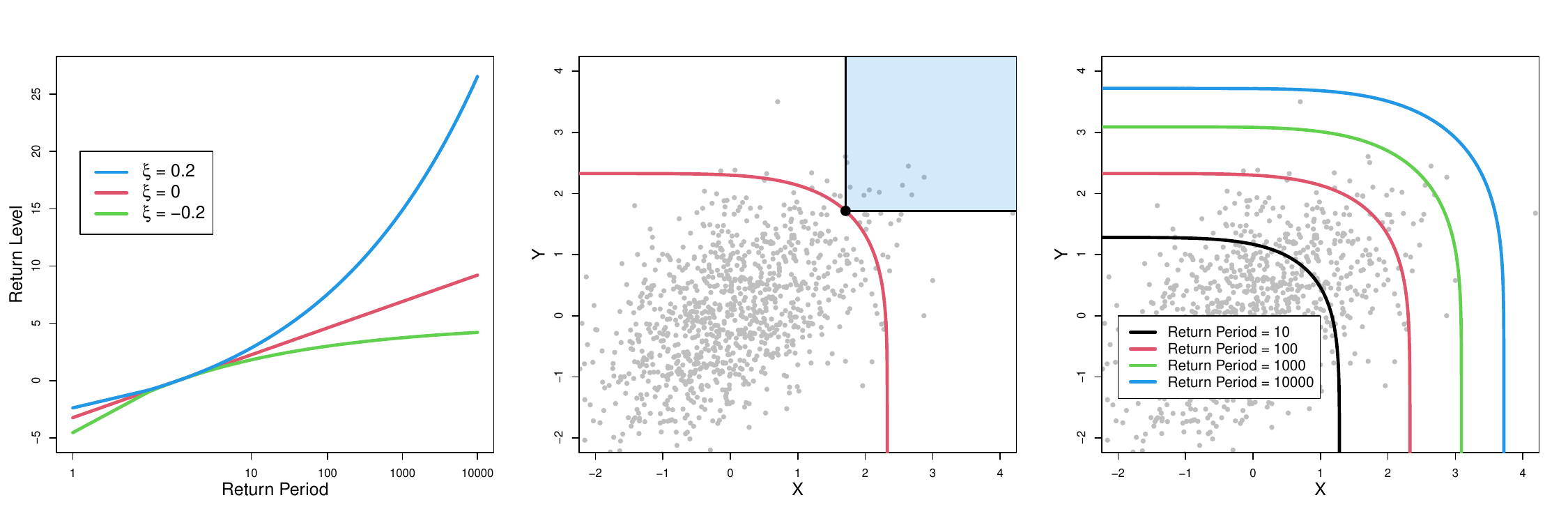}
    \caption{Left: Return level plots under varying shape parameters. Centre: Return curve $\RC{p}{}$ (red) of standard bivariate normal data with $\rho = 0.5$ and $n=1000$ for return period of 100 ($p = 1/100$). In a sample of size $n$, we expect to observe $np$ points in the blue shaded region; this is true for all equivalent shaded regions at any point on the curve. Right: Return curves of the same bivariate normal data set for return periods in the set $\{10,100,1000,10000\}$.}
    \label{fig:return_level_plots}
\end{figure}

Often, univariate extreme value models are used to estimate risk measures for events associated with small probabilities; these summary statistics provide a quantification of extremal behaviour that can be used to help mitigate against rare environmental events, such as floods, storms, or wildfires. One such measure is known as a return level. Given a real variable $X$, representing a measurement taken at regular time intervals, and probability $p$, the $p$-probability return level is the value $x_p$ that satisfies the equation $\Pr(X>x_p) = p$. For small $p$, $x_p$ represents a high quantile which can be estimated using the GPD. We restrict attention to the case when $X$ is stationary since the interpretation of return levels is more straightforward in this setting. We define the return period of $x_p$ to be the value $1/p$; one would expect the variable $X$ to exceed $x_p$ once, on average, during each return period. The relationship between return levels and periods can be illustrated using a return level plot; examples of three such plots with varying shape parameters are given in the left panel of Figure \ref{fig:return_level_plots}.

Return levels are widely used and provide a simple way to understand risk. However, many potentially impactful events arise due to the effect of more than one variable. For example, \citet{Mattei2001} describe how the combination of high sea levels and wind led to large-scale flooding at the Blayais nuclear power plant in 1999, causing significant damage. For this reason, it is desirable to have similar risk measures in the multivariate case, but thus far relatively little consideration has been given to this problem. This is in part due to the lack of natural ordering for multivariate vectors, which means there is no longer a single definition of an extreme event. A variety of measures have been proposed \citep{Serinaldi2015}, with each suited to a different analytical need. In this text, we focus on the bivariate case and one particular measure known as a return curve, which directly extends the return level concept to the bivariate setting.

\subsection{Return curves}
Consider the joint survival function of the continuous random vector $(X,Y)$ at a given probability $p$, i.e. $\Pr(X>x,Y>y) = p$. The combinations $(x,y) \in \RR^2$ satisfying this equation define a curve in the plane; we therefore define the $p$-probability return curve to be the set 
\begin{equation*}
    \RC{p}{} := \left\{ (x,y) \in \RR^2: \Pr(X>x,Y>y) = p \right\}.
\end{equation*} 
We consider values of $p$ close to zero, corresponding to rare joint exceedance events. Within the literature, this set has a variety of labels, including isolines \citep{Cooley2019}, hazard curves \citep{Simpson2017} and joint probability curves \citep{Gouldby2017}. In an analogue to return levels, we define the return period to be $1/p$, since given any point $(x,y) \in \RC{p}{}$, we would expect to observe the event $\{X>x,Y>y\}$ once, on average, each return period. Equivalently, in a sample of size $n$ from $(X,Y)$, we expect to observe $np$ points in the region $(x,\infty) \times (y,\infty)$. 

Since return curves define a line in $\RR^2$ rather than a single value, the two dimensional return level plot does not naturally extend to this setting. Instead, we can consider different return periods and plot the corresponding curves individually or simultaneously; examples of both are given in the centre and right panels of Figure \ref{fig:return_level_plots} for a standard bivariate normal data set with correlation coefficient $\rho = 0.5$. 


Return curves are arguably the most intuitive bivariate extension to return levels since they are also defined in terms of the survivor function. They have been used in practice to derive extremal environmental conditions for the design and analysis of ocean structures, such as oil rigs \citep{Jonathan2014}, freight ships \citep{Vanem2020} and wind turbines \citep{Manuel2018,Velarde2019}, and coastal structures, such as railway lines \citep{Agency2005,Gouldby2017} and wave energy converters \citep{Eckert-Gallup2016}. 

\begin{figure}[!ht]
    \centering
    \includegraphics[width=\linewidth]{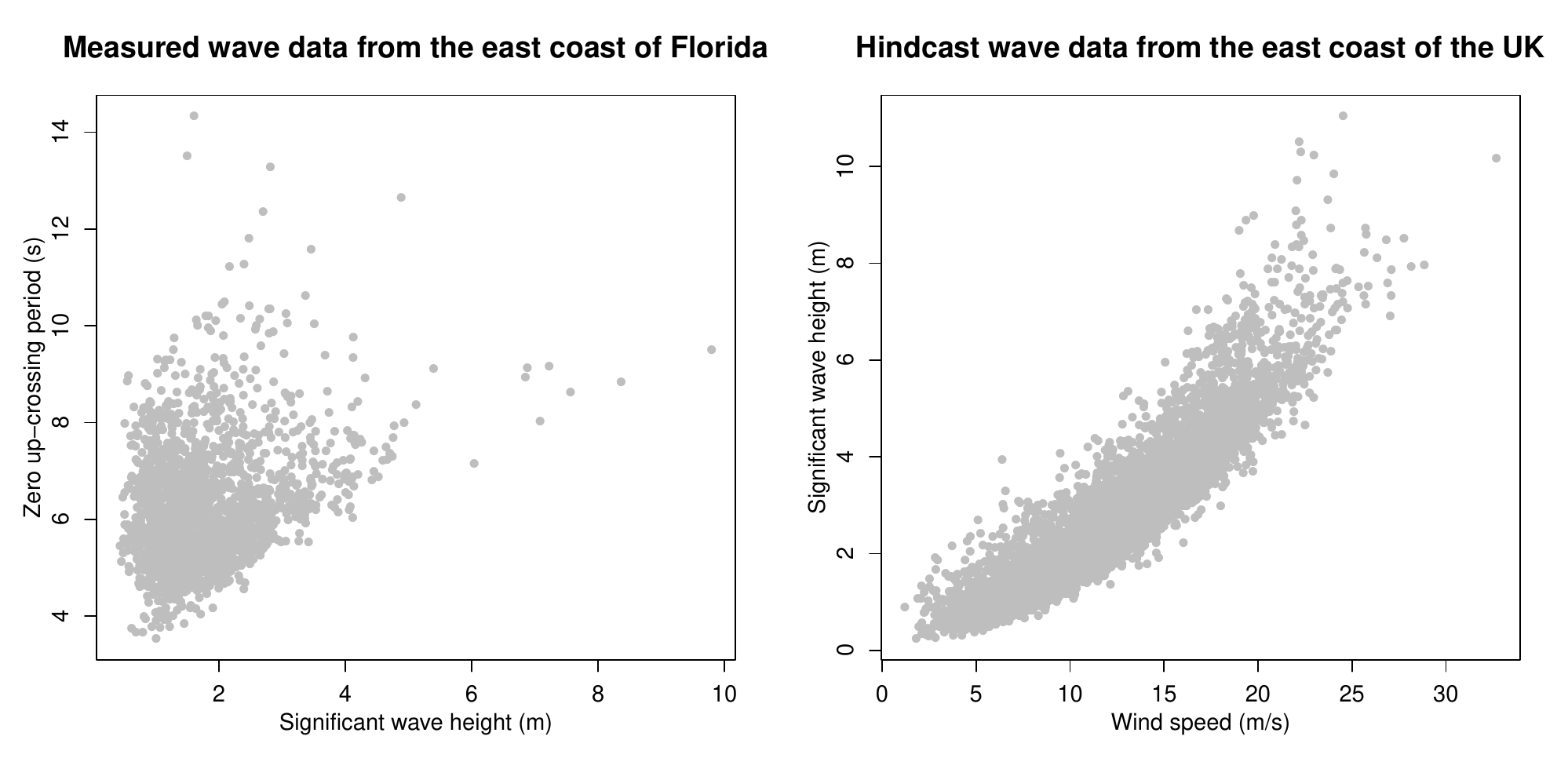}
    \caption{Left: Measured metocean data for a buoy located on the east coast of Florida, USA.  Right: Hindcast metocean data for a location near the east coast of the UK.}
    \label{fig:env_data}
\end{figure}

As motivating examples, we consider two environmental data sets of practical importance, both of which are illustrated in Figure \ref{fig:env_data}. Our objective is to use return curve estimates to derive joint extremal conditions for each data set. Both data sets are comprised of metocean variables, which have previously been used in a comparison exercise for a risk measure known as an environmental contour \citep{Haselsteiner2021}; such measures also aim to summarise joint extremal behaviour. However, unlike return curves, they do not offer an intuitive interpretation in terms of return periods. 

The first data set contains measured significant wave height (m) and zero up crossing period (s) between 1996-2005 obtained from a buoy on the east coast of Florida, USA. The second data set contains 25 years of wind speed (m/s) and significant wave height (m) observations obtained from the hindcast model coastDat $-2$ \citep{Groll2017} for a location in the North Sea near the east coast of the UK. These combinations of variables are of particular relevance for the structural reliability of offshore and coastal structures, and bivariate risk measures are commonly used to inform the design basis for such structures \citep{Jonathan2014,Haselsteiner2019,Mackay2020}. They therefore provide realistic examples with which to illustrate the utility of return curve estimates.  



Both of the original data sets are comprised of hourly observations, resulting in strong temporal dependence between observations. To reduce the temporal dependence while retaining a relevant data set, we restrict attention to daily maxima values. Furthermore, we account for seasonal non-stationarity by restricting attention to the months September-March; our analysis suggests the largest wave heights are observed in this interval. This results in $n=2048$ and $n=5306$ observations, respectively, for the measured and hindcast data sets.

\subsection{Practical considerations for return curve estimation}

To estimate return curves in practice, we must first estimate the joint survival function of an observed bivariate process at a fixed probability $p$. Since we restrict attention to probabilities close to zero, we require an estimation method that can accurately capture the joint tail behaviour of the process and provide realistic extrapolation to estimate $\RC{p}{}$ for $p$ outside of the observation period. For this reason, we restrict attention to models from multivariate extreme value theory. We must also consider methods for evaluating the uncertainty associated with return curve estimates. This is a more complex problem than assessing the uncertainty of return level estimates, since there is more than one dimension in which the risk measure can vary. Furthermore, given a data set for which a curve has been estimated at probability $p$, it is also essential to evaluate goodness of fit via a diagnostic tool. Little attention has been given to either of these problems within the extremes literature, motivating the development of novel approaches. 

This article is structured as follows. Section \ref{Sec2} provides a brief background on bivariate extreme value theory. In Section \ref{Sec3}, we introduce various properties of return curves, review existing approaches for curve estimation within the extremes literature and present two novel approaches. Section \ref{Sec4} introduces new tools for analysing uncertainty and goodness of fit in return curve estimates. Section \ref{Sec5} presents a simulation study to evaluate the performance of these tools and compare curve estimates from several models. In Section \ref{Sec6}, we apply our methodology to the two aforementioned environmental data sets. We conclude in Section \ref{Sec7} with a discussion and outlook on future work. 

\section{Bivariate extreme value theory} \label{Sec2}

When assessing the extremal behaviour of a continuous bivariate random vector $(X,Y)$, one must consider tail behaviour within both marginal processes, along with the dependence between the largest observations of either variable, which we refer to as the extremal dependence. A fundamental classification of extremal dependence is whether the most extreme events can occur jointly or not. This is quantified by the coefficient $\chi:=\lim_{u \to 1}\chi(u) \in [0,1]$, where 
\begin{equation*}
    \chi(u) = \Pr(F_Y(Y) > u \mid F_X(X) > u),
\end{equation*}
for $X \sim F_X$, $Y \sim F_Y$. The cases $\chi = 0$ and $\chi > 0$ correspond to asymptotic independence (AI) and asymptotic dependence (AD), respectively. A larger $\chi$ corresponds to stronger positive dependence in the joint tail, with $\chi=1$ occurring for perfectly dependent variables. In practice, we cannot estimate $\chi$ in the limit as $u \to 1$, meaning we must use $\chi(u)$, $u < 1$, to determine AD vs AI. 

Early modelling techniques proposed for multivariate extreme values focused only on the AD case \citep{DeHaan1977,Coles1991,DeHaan1998}. Such methods are based on the framework of multivariate regular variation: given a random vector $(X,Y)$ with standard Fr\'echet margins, we define radial and angular components $R:= X+Y$ and $W := X/R$, respectively. We say $(X,Y)$ is multivariate regularly varying if, for measurable $B \subset [0,1]$, we have 
\begin{equation} \label{eqn:MRV}
    \lim_{r \to \infty} \Pr (W \in B, R > sr \mid R>r) = H(B) s^{-1}, \; s \geq 1,
\end{equation}
with $H(\partial B) = 0$, where $\partial B$ is the boundary of $B$ \citep{Resnick1987}. Assumption \eqref{eqn:MRV} implies that for large radial values, $R$ and $W$ are independent. The quantity $H$, which is known as the spectral measure, captures the extremal dependence structure of $(X,Y)$ and must satisfy the moment constraint $\int_0^1wH(\mathrm{d}w) = 1/2$. For all AI vectors, the spectral measure places mass on the atoms $\{0\}$ and $\{1\}$; as such, this modelling framework cannot capture tail properties under this extremal dependence scheme \citep{Coles1999}.

In recent years, it has been shown that the AI case is at least as important as the AD case, and that assuming the incorrect form of extremal dependence leads to unsuitable extrapolation in the joint upper tail \citep{Ledford1996,Ledford1997,Heffernan2004}. Therefore, unless there is strong prior knowledge in favour of either AD or AI, it is desirable to use models which have sufficient flexibility to allow the data to directly inform the class of extremal dependence structure. 

The first such approach was proposed in \citet{Ledford1996}. Given $(X,Y)$ with standard exponential margins, this model assumes the joint tail representation
\begin{equation} \label{eqn:led_tawn}
    \Pr (X > t, Y > t) = \Pr(\min(X,Y)>t) = L(e^t)e^{-t/\eta} \hspace{.5em} \text{as} \; t \to \infty,
\end{equation}
where $L$ is a slowly varying function at infinity, i.e., $\lim_{t \to \infty}L(ct)/L(t) = 1$ for $c>0$, and $\eta \in (0,1].$ The parameter $\eta$ is termed the coefficient of tail dependence, with $\eta=1$ and $\lim_{t \to \infty}L(t) > 0$ corresponding to AD and $\eta < 1$, or $\eta = 1$ and $\lim_{t\to \infty}L(t) = 0$, corresponding to AI. Estimation of $\eta$ can be performed in practice using the Hill estimator \citep{Hill1975}. Several extensions to this approach exist \citep{Ledford1997,Ramos2009}: however, all techniques derived under this framework are applicable only within regions where both variables are large. Consequently, these methods are not appropriate for the estimation of return curves, since this measure is defined also in regions where only one variable is large; see Figure \ref{fig:return_level_plots}. 

\citet{Wadsworth2013} provide an alternative representation for bivariate tail probabilities using a more general extension of the model described in equation \eqref{eqn:led_tawn} that allows for joint tail estimation in regions where only one variable is large. Given $(X,Y)$ with standard exponential margins, they assume for each $w \in [0,1]$
\begin{equation} \label{eqn:wads_tawn}
    \Pr(\min\{X/w,Y/(1-w)\}>t) = L(e^t\mid w)e^{-\lambda(w)t}, \; \lambda(w) \geq \max(w,1-w),
\end{equation}
as $t \to \infty$, where $L(\cdot \mid w)$ is slowly varying for each ray $w \in [0,1]$. The function $\lambda$, which is termed the angular dependence function, is the key quantity in determining joint tail behaviour, and both AD and AI can be captured under this assumption, with AD implying the lower bound $\lambda(w) = \max(w,1-w)$. This quantity generalises the coefficient $\eta$, with $\eta = 1/(2\lambda(0.5))$, and can be estimated pointwise for any ray $w \in [0,1]$ using the Hill estimator. This approach can be used to estimate joint survivor probabilities where only one variable is large by taking values of $w$ close to $0$ or $1$. 


\citet{Heffernan2004} proposed a very general modelling tool for conditional probabilities. We consider the extension given in \citet{Keef2013} since the formulation given in the original approach cannot easily accommodate structures exhibiting negative dependence. Given a random vector $(X,Y)$ with standard Laplace margins, it is assumed that there exist normalising functions $a:\RR \to \RR$ and $b:\RR \to \RR_+$ such that
\begin{equation} \label{eqn:heff_tawn}
    \lim_{t \to \infty}\Pr\left[(Y-a(X))/b(X) \leq z, X-t > x \mid X>t\right] = D(z)e^{-x},
\end{equation}
for a non-degenerate distribution function $D$. Similarly to \citet{Wadsworth2013}, this framework is able to capture both AD and AI, with AD arising when $a(x) = x$ and $b(x) = 1$. This method is a flexible approach for modelling multivariate extremes and is also not restricted only to regions where both variables are large. Note that one could also condition on the event $Y>t$ and assume the existence of normalising functions for the variable $X$: in combination, these assumptions allow consideration of the region where either variable is large. The functions $a$ and $b$ are typically estimated parametrically under a misspecified model for $D$, while the distribution function $D$ is subsequently estimated non-parametrically. 

Alongside these approaches, we note that there exist a range of copula-based models that can capture both dependence regimes \citep{Coles2002,Wadsworth2017,Huser2019}. Such techniques aim to create a unified modelling framework for AD and AI. Moreover, the case of AD does not represent a boundary case for the latter two approaches, which could be practically advantageous. However, they all require stronger assumptions about the form of parametric family for the bivariate distribution, reducing their flexibility and limiting their use in practice. As a result, we prefer instead to consider the more flexible modelling assumptions described in equations \eqref{eqn:wads_tawn} and \eqref{eqn:heff_tawn}.

\section{Bivariate return curve estimation} \label{Sec3}
We now consider techniques for practical estimation of return curves. We begin by introducing theoretical results for return curves in Section \ref{subsec:RC_theory}. Naive implementation of statistical estimation methods would generally produce curves that fail to respect these results, but imposing them will typically improve estimation. Section \ref{subsec:marg_trans} details how to perform marginal transformations so that the models in equations \eqref{eqn:wads_tawn} and \eqref{eqn:heff_tawn} can be applied in practice. In Section \ref{subsec:existing_return_curve}, we present existing approaches for return curve estimation before introducing novel estimation techniques in Section \ref{subsec:novelRCestimation}.
\subsection{Return curve properties} \label{subsec:RC_theory}
A careful consideration of the theory surrounding the joint survival function allows us to deduce several properties about the shape and magnitude of $\RC{p}{}$ for a given $p \in (0,1)$. We begin by noting that the joint distribution function, $F_{X,Y}$, can be expressed in terms of the marginal distribution functions of $X$ and $Y$, $F_X$, $F_Y$, respectively, and a copula $C$ via $F_{X,Y}(x,y) = C(F_X(x),F_Y(y))$. The return curve is linked to the joint distribution function $F_{X,Y}$ by the equation $\Pr(X>x,Y>y) = 1 - F_X(x) - F_Y(y) + F_{X,Y}(x,y)$. Throughout this section, we make the assumption that the random vector $(X,Y)$ has strictly continuous marginal distribution functions. 
\begin{property} \label{prop:marg_limit}
Let $x_p:=F_{X}^{-1}(1-p)$ and $y_p:=F_{Y}^{-1}(1-p)$ be the $(1-p)$-th quantiles of X and Y, respectively. Then for $(x,y) \in \RC{p}{}$, $x \leq x_p$ and $y \leq y_p$.
\end{property}
\begin{proof}
We have $\Pr(X>x) \geq \Pr(X>x,Y>y) = p = \Pr(X>x_p)$ and hence $x \leq x_p$ .

\end{proof}
This result bounds the coordinate values that can be observed on the return curve. Next, by considering the limit of the joint survivor function as one variable converges to the lower limit of the marginal support, we obtain the following result. \begin{property} \label{prop:inf_supp}
    Let $\text{supp}(F)$ denote the support of $F$ and $x_{inf} := \inf \{\text{supp}(F_X)\}$, $y_{inf} := \inf \{\text{supp}(F_Y)\}$. We have that 
    \begin{equation*}
    \Pr(X>x,Y>y) = \begin{cases}
     \Pr(Y>y) \; &\text{if} \; x\leq x_{inf}, \\
    \Pr(X>x) \; &\text{if} \; y\leq y_{inf}.
    \end{cases}
    \end{equation*}
\end{property} 
Combining this statement with Property \ref{prop:marg_limit}, Property \ref{prop:curve_limits} follows. 
\begin{property} \label{prop:curve_limits}
Let $(x,y) \in \RC{p}{}$. If $x\leq x_{inf}$ $(y\leq y_{inf})$, then $y = y_p$ $(x=x_p)$.  
\end{property}

These results allow us to easily compute the curve coordinates on the regions $(-\infty, x_{inf}) \times (y_p, \infty)$ and $(x_p, \infty) \times (-\infty, y_{inf})$, assuming we can accurately estimate the marginal quantiles $(x_p, y_p)$ and the infima of marginal supports $(x_{inf}, y_{inf})$. Finally, by considering coordinates at different points on a return curve, we obtain the following result.
\begin{property} \label{prop:shape_curve}
Suppose the copula, C, of $(X,Y)$ on uniform margins has joint support on the whole of $[0,1]^2$ and joint density function, denoted $c$.  Given $(x_1,y_1),(x_2,y_2)\in\RC{p}{}$ with $0<F_X(x_1),F_X(x_2)<1$ and $0<F_Y(y_1),F_Y(y_2)<1$, we have that $x_1 < x_2 \Leftrightarrow y_1 > y_2$. 
\end{property}
\begin{proof}
Suppose $x_1 < x_2$ and $y_1 \leq y_2$. This implies that
\begin{align*}
    p &= \Pr(X>x_1,Y>y_1) = \Pr(F_X(X)>F_X(x_1),F_Y(Y)>F_Y(y_1)) \\
    &= \int_{F_X(x_1)}^1 \int_{F_Y(y_1)}^1 c(u,v) \mathrm{d}v \mathrm{d}u \\
    &= \int_{F_X(x_1)}^{F_X(x_2)} \int_{F_Y(y_1)}^1 c(u,v) \mathrm{d}v \mathrm{d}u + \int_{F_X(x_2)}^1 \int_{F_Y(y_1)}^1 c(u,v) \mathrm{d}v \mathrm{d}u \\
    &>  \int_{F_X(x_2)}^1 \int_{F_Y(y_1)}^1 c(u,v) \mathrm{d}v \mathrm{d}u \hspace{1em} (\text{since we have support on the whole of $[0,1]^2$}) \\
    &\geq \int_{F_X(x_2)}^1 \int_{F_Y(y_2)}^1 c(u,v) \mathrm{d}v \mathrm{d}u = \Pr(X>x_2,Y>y_2) = p
\end{align*}
implying $p>p$, a contradiction. Hence, $y_1 > y_2$. 
\end{proof}
This result governs the shape of the contour defined by the return curve set. We note there is an alternative proof given in \citet{Cooley2019} under the assumption of monotonicity of the joint survivor function.  

\subsection{Marginal transformations} \label{subsec:marg_trans}
From Section \ref{Sec2}, it is clear that in order to apply multivariate extreme value models, we need to standardise the marginal distributions of a random vector to achieve the form assumed by the model. Typically, inference involves two steps: forward transformation to get the data onto desired margins and back transformation to move any computed statistics, such as a return curve, back onto the original margins. For both steps, we use the semi-parametric approach given in \citet{Coles1991}. Given an identically distributed sample $\{(x_i,y_i):i=1,\hdots,n\}$ from a random vector $(X,Y)$ with unknown margins, we estimate the marginal distribution $\hat{F}_X$ (similarly $\hat{F}_Y$) by 
\begin{equation} \label{eqn:CDF_trans}
    \hat{F}_X(x) = \begin{cases} 1 - \{1 - \Tilde{F}_X(u_{X})\}\{1 + \xi_X(x-u_{X})/\sigma_X \}_+^{-1/\xi_X}, \hspace{1em} &\text{for} \; x > u_{X}, \\  \Tilde{F}_X(x), \hspace{1em}  &\text{for} \; x \leq u_{X},  \end{cases}
\end{equation}
where the first line represents the GPD above a high threshold $u_X$ and $\Tilde{F}_X$ is an empirical rank transform given by 
$\Tilde{F}_X(x) = \sum_{i=1}^n \mathbbm{1}(x_{i} \leq x)/(n+1)$. This approach ensures the marginal tail behaviour is captured within the transformation. Moreover, equation \eqref{eqn:CDF_trans} can be easily inverted to perform the back transformation step. 

\subsection{Existing methodology} \label{subsec:existing_return_curve}

The literature on extremal return curve estimation is sparse, owing to the fact that little consideration has been given to this problem in practice. Of the existing approaches, each can be designated into one of three categories: approaches for AD data only, approaches for AI data only, and approaches applicable to data exhibiting either regime. The majority of the available literature falls within the first of these categories \citep{Stephenson2002,Salvadori2004,Marcon2017}. In all cases, the authors assume the bivariate copula is in a family of distributions termed bivariate extreme value copulas. These copulas, which are directly related to the spectral measure in equation \eqref{eqn:MRV}, imply AD and provide the basis for the majority of multivariate extreme value techniques. The necessity for making such a strict assumption is a well known drawback of this kind of model, since the form of extremal dependence is seldom known prior to analysis and AI is frequently observed in practice \citep{Heffernan2004,Huser2019}. We therefore choose not to consider such approaches further. 

In \citet{Cooley2019}, the authors propose separate techniques for the first and second categories. In both cases, extremal return curve estimates are obtained by `projecting' empirical curves estimated for less extreme probabilities. For the case of AD, given a random vector $(X,Y)$ with standard Fr\'echet margins, multivariate regular variation is exploited to obtain curve estimates. Consider two small probabilities $p$ and $p^*$ with $p^*>p$; multivariate regular variation implies that $\RC{p}{} \approx s^{-1}\RC{p^*}{}$, where $s:=p^*/p>1$. In practice, $\RC{p^*}{}$ is estimated empirically via a smooth Gaussian-kernel estimate of the joint survivor function, and scaled by the coefficient $s^{-1}$ to produce an estimate for $\RC{p}{}$. 

In the case of AI, a similar estimation procedure is proposed based on the framework of hidden regular variation \citep{Resnick2002}, an elaboration of the assumption outlined in equation \eqref{eqn:led_tawn}. However, as mentioned in Section \ref{Sec2}, this approach only works in regions where both variables are large. To account for this, \citet{Cooley2019} proposed an ad-hoc procedure to link this region to the marginal axes, requiring additional steps and parameter estimation. In contrast, the assumption in equation \eqref{eqn:wads_tawn} provides a theoretically sound link between regions where variables are of different magnitudes. Estimates of return curves from the \citet{Cooley2019} approach are illustrated in Figure \ref{fig:copula_examples}. In Section \ref{Sec5}, we compare the resulting curve estimates from this approach to the methods introduced in this paper and show that the techniques we present outperform this method in a wide range of scenarios. 

For the third category, few approaches exist within the literature; this is in part because the bivariate extreme value methodologies that allow for dual estimation are relatively modern. All proposed techniques use a semi-parametric implementation of the conditional extremes model described in equation \eqref{eqn:heff_tawn} \citep{Jonathan2014,Gouldby2017,Simpson2017}. However, like the other techniques introduced in this section, little to no consideration is given to the theory behind return curves, leading to curve estimates with undesirable properties. Moreover, as will be discussed in Section \ref{subsubsec:HTmethod}, utilising the \citet{Heffernan2004} modelling framework for return curve estimation is not straightforward, requiring delicate treatment and several steps; this has not been fully acknowledged in these existing approaches. 

To the best of our knowledge, the modelling techniques discussed here cover all of the proposed methods for estimating return curves at extremal probabilities. Furthermore, we know of no attempt to compare curve estimates from these different methods. No formal quantifications of return curve uncertainty or bias have been proposed previously, making it difficult to evaluate performance over different dependence structures. Some approaches \citep{Simpson2017,Cooley2019} instead provide bootstrap curve estimates which, while representing the uncertainty in curve estimates, do not provide interpretable confidence regions.

Alongside this issue, there is only one diagnostic tool in the literature for evaluating the accuracy of return curve estimates \citep{Cooley2019}. This tool utilised the result that, if data are independent and identically distributed, the number of points in each survival region on the return curve should theoretically be Binomial$(n,p)$ distributed, where $n$ denotes the size of the data set. This property can be used to construct a theoretical confidence region for the number of points in the joint survival set $(x,\infty) \times (y,\infty)$ for any point $(x,y) \in \RC{p}{}$. While the authors show that these confidence regions capture the number of observations within estimated joint survival sets for the majority of considered examples, we argue that the resulting diagnostic is relatively uninformative since it lacks an intuitive interpretation in terms of the survival probability $p$. Moreover, this diagnostic strongly relies on the assumption of independent observations, which is seldom the case in practice. We present an alternative diagnostic tool in Section \ref{subsec:diagnostic}, where confidence intervals are instead obtained using sets of empirical probability estimates obtained through bootstrapping and compared to the true probability $p$. Temporal dependence can be incorporated through block bootstrapping, meaning this tool can be applied to a wider range of data sets; see Sections \ref{subsec:diagnostic} and \ref{Sec6} for further details.

\subsection{Novel methods for return curve estimation}
\label{subsec:novelRCestimation}

We outline two methods for estimation of $\RC{p}{}$ based on the modelling assumptions given in equations \eqref{eqn:wads_tawn} and \eqref{eqn:heff_tawn}. Consider a random vector $(X,Y)$ with standard exponential margins, for which the marginal support is given by the set $\RR_+$; this implies $x_{inf}=y_{inf}=0$. We can immediately deduce from Property \ref{prop:curve_limits} that the coordinates of the return curve intersecting the margins are given by $(0,y_p)$ and $(x_p,0)$, with $y_p = x_p = -\log(p)$, the $(1-p)$-th quantile. These coordinates give us `start' and `end' points for curve construction. Moreover, given a curve estimate $\widehat{\mathrm{RC}}(p)$ for this vector constructed with these boundary points, Properties \ref{prop:marg_limit} and \ref{prop:shape_curve} can be imposed to ensure the resulting curve is theoretically possible. For the former, if any $(x,y)\in \widehat{\mathrm{RC}}(p)$ satisfy $x>x_p$ (similarly, $y>y_p$), we set $x = x_p$ ($y=y_p$), thereby bounding values on the curve. For the latter, we treat the bounded curve estimate as a function of $x$ and apply an iterative algorithm starting at the point $(0,y_p)$ to obtain a monotonic function. Incorporating additional theoretical knowledge into return curve estimation should lead to more accurate and robust estimates. For both of the methods introduced in this section, we impose the properties above retrospectively once curve estimates have been obtained. 

\subsubsection{Method based on the approach given in \texorpdfstring{\citet{Heffernan2004}}{TEXT}}\label{subsubsec:HTmethod}

In this section, we propose an implementation of the \citet{Heffernan2004} model, which builds on the existing methods that have applied this framework for return curve construction. Unlike these techniques, we incorporate the properties introduced in Section \ref{subsec:RC_theory} into return curve estimates and provide an intuitive algorithm for combining the point estimates obtained from conditioning on both variables. Let $(X_L,Y_L)$ denote the vector $(X,Y)$ on standard Laplace margins and consider a small probability $p$ for which we wish to obtain a return curve estimate. To achieve this, we fit the \citet{Heffernan2004} model twice, conditioning on both $X_L$ and $Y_L$ separately, thus allowing us to estimate the curve in different regions. In particular, we consider the regions defined by $ R_{Y_L>X_L}:=\{ (x_L,y_L) \in \RR^2 \mid y_L>x_L\}$ and $R_{Y_L\leq X_L}:=\{ (x_L,y_L) \in \RR^2 \mid y_L\leq x_L\}$.

For $R_{Y_L>X_L}$, we first select a high quantile $u_{Y_L}$ from the marginal distribution of $Y_L$ such that $\Pr(Y_L>u_{Y_L}) > p$. In particular, we select the $0.95$ empirical quantile of this distribution, implying the return curve probability, $p$, must be smaller than $0.05$. We assume the normalising functions are given by $a(y) = \alpha y$ and $b(y) = y^{\beta}$ for constants $\alpha \in [-1,1]$ and $\beta \in (-\infty,1)$: as noted in \citet{Keef2013}, these functions capture the limiting dependence structures for a wide range of distributions. The parameters $\alpha$ and $\beta$ can be estimated under the working assumption that the distribution function $D$, which captures the stochastic behaviour of the variable $(X_L - \alpha Y_L)/Y_L^{\beta} \mid Y_L>u_{Y_L}$, is that of a Gaussian distribution. We denote the fitted values by $\hat{\alpha}$ and $\hat{\beta}$. These values can be used to simulate from the variable $X_L \mid Y_L>u_{Y_L}$; for example, see \citet{Jonathan2014}.

We then consider a decreasing sequence of high quantiles from $Y_L$ that exist in the interval $(u_{Y_L}, F_{Y_L}^{-1}(1-p))$. The upper end point of this interval is the limit that $Y_L$ values can attain on this curve and the lower end point represents the minimal quantile for which the fitted model is valid. We denote this set by $\mathcal{Y}$ and iteratively consider each $y_* \in \mathcal{Y}$ in turn, with $q := \Pr(Y_L>y_*)$. Using the fitted parameter values, we use the model to simulate from the conditional distribution $X_L \mid Y_L > y_*$. Letting $x_*$ denote the (estimated) $(1-p/q)$-th quantile from this distribution, we have that the resulting coordinate $(x_*,y_*)$ is a member of the set $\widehat{\mathrm{RC}}(p)$ (defined for $(X_L,Y_L)$) since $\Pr(X_L>x_*,Y_L>y_*) = \Pr(X_L>x_* \mid Y_L>y_*)\Pr(Y_L>y_*) = \frac{p}{q}\times q = p$. We continue in this manner until we obtain a value $x_{**}$ with $y_* \leq x_{**}$, or we have exhausted all values in the set $\mathcal{Y}$. The resulting coordinate set then gives an estimate of the curve in $R_{Y_L>X_L}$. 

\begin{figure}[!tb]
    \centering
    \includegraphics[width=.5\textwidth]{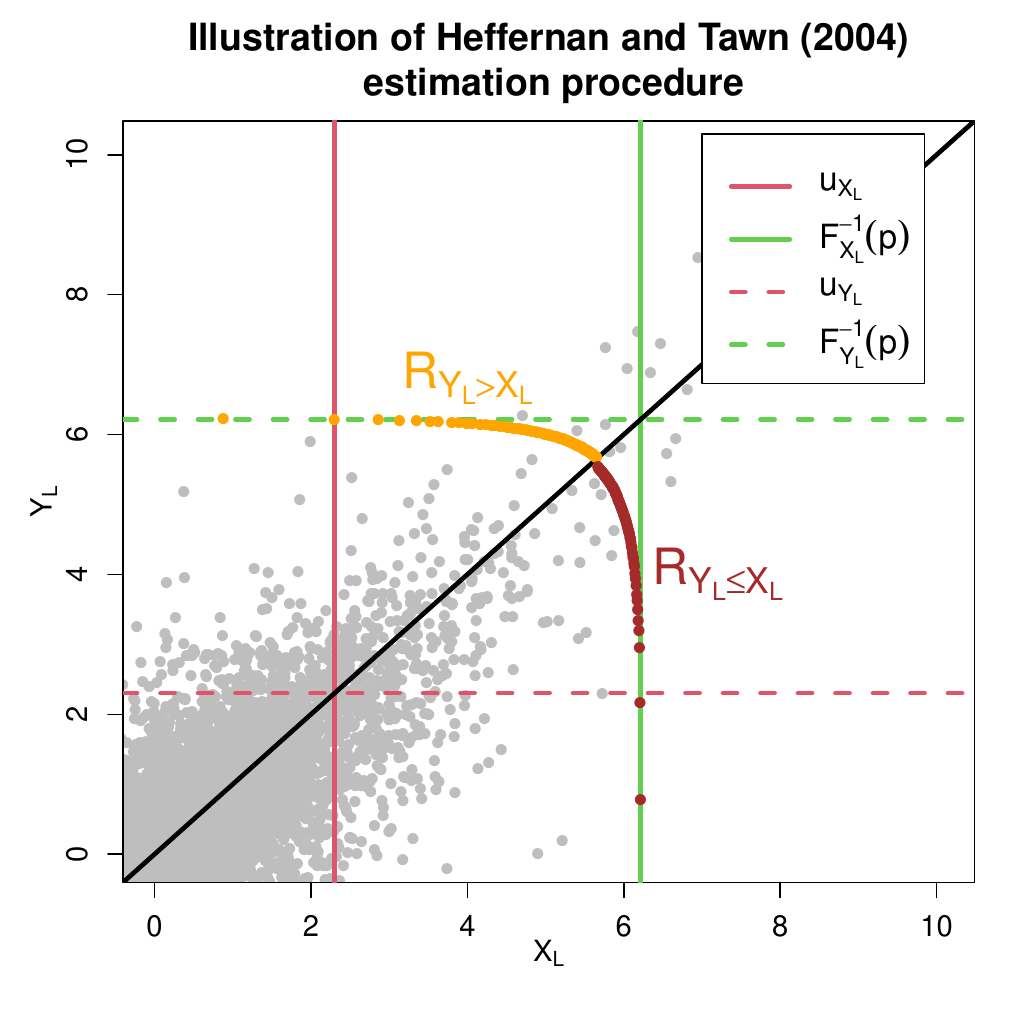}
    \caption{Illustration of return curve estimation procedure using \citet{Heffernan2004} modelling framework. Red and green lines give 0.95 and $(1-p)$-th quantiles for both margins, respectively. The brown and orange points denote the point estimates obtained for the regions $R_{Y_L>X_L}$ and $R_{Y_L\leq X_L}$, respectively.}
    \label{fig:heff_tawn_procedure}
\end{figure}

A near identical procedure is used to obtain the curve estimate in $R_{Y_L\leq X_L}$, this time selecting a high quantile $u_{X_L}$ from the distribution of $X_L$ and fitting the conditional model above this quantile. We then consider a set of quantiles in the interval $(x^{'}_{**},F^{-1}_{X_L}(1-p))$, where $x^{'}_{**} = x_{**}$ if $x_{**}$ exists and $u_{X_L}$ otherwise, where $u_{X_L}$ denotes the empirical $0.95$ quantile from $X_L$. We label this set $\mathcal{X}$, ordered such that the quantiles are increasing, and use the fitted model to obtain quantiles from the conditional distribution $Y_L \mid X_L > x_*$ for each $x_* \in \mathcal{X}$. The resulting coordinate sets from both regions are combined to give an estimate of the return curve over the entire joint support of $(X_L,Y_L)$. An illustration of this procedure is given in Figure \ref{fig:heff_tawn_procedure}. As can be observed, two sets of points estimates are obtained by conditioning on either variable; these sets are then combined to give an estimate of the entire return curve. As a final step, we apply the probability integral transform to transform the curve estimate to standard exponential margins. 

We note that the implementation of this model to estimate return curves is more complex than the methods proposed in Section \ref{subsubsec:WTmethod} and \citet{Cooley2019}. This is due to the fact the model requires a variable to condition on, meaning we have to fit the model twice and provide a technique for joining point estimates from regions $R_{Y_L>X_L}$ and $R_{Y_L\leq X_L}$. 

\subsubsection{Method based on the approach given in \texorpdfstring{\citet{Wadsworth2013}}{TEXT}}\label{subsubsec:WTmethod}

In this section, we propose a novel implementation of the model described in equation \eqref{eqn:wads_tawn} to generate non-parametric return curve estimates. Consider a random vector $(X,Y)$ with standard exponential margins and define a set $\mathcal{W}$ containing equally spaced rays in the interval $[0,1]$, ordered from lowest to highest. Assuming $\vert \mathcal{W} \vert$, is sufficiently large, we are able to evaluate the joint extremal behaviour across the entire region for which at least one variable is extreme. For each $w \in \mathcal{W}$, we use the $95\%$ empirical threshold of the variable $T_w := \min\left\{ X/w, Y/(1-w) \right\}$ to obtain an estimate of the angular dependence function via the Hill estimator, which we denote $\hat{\lambda}(w)$. For large $u$, equation \eqref{eqn:wads_tawn} implies that for any $w \in (0,1)$ and $t>0$, 
\begin{equation*}
    \Pr\left( T_w> t + u \Big\vert T_w > u \right) \approx \exp\{-t\hat{\lambda}(w)\}.
\end{equation*}
Estimates of $t$ and $u$, combined with the rays $w$, provide estimates of points in $\RC{p}{}$. We firstly select a small probability $p^*>p$ and estimate $u$ as the $(1-p^*)$-th quantile of $T_w$, implying $\Pr(T_w >u) = p^*$. One can then estimate the value of $t>0$ such that $\Pr(T_w >t+u) = p$ since
\begin{align*}
    p = \Pr(T_w >t+u)= \Pr(T_w >u) \times \Pr(T_w >t+u \mid T_w >u)  = p^*\exp\{-t\hat{\lambda}(w)\}, 
\end{align*}
giving the estimate $t = -\frac{1}{\hat{\lambda}(w)}\log(p/p^*)$. Setting $(x,y) := (w(t+u),(1-w)(t+u))$, we have $(x,y) \in \widehat{\mathrm{RC}}(p)$, resulting in a return curve point estimate for each ray $w \in \mathcal{W}$.

\section{Uncertainty estimation and diagnostic tool} \label{Sec4}
\subsection{Capturing uncertainty in return curve estimates}\label{subsec:quantifying_uncertainty}
While previous methods for return curve estimation have considered sampling uncertainty, none provide a means to construct interpretable confidence regions and/or `average' estimates for return curves \citep{Simpson2017,Cooley2019}. Here, we propose a new method for representing uncertainty in return curve estimates that addresses limitations in the existing methods and provides a formal framework for comparing curve estimates from different models where the truth is known. 


Our goal is to represent sampling uncertainty in return curve estimates via some type of confidence region at a given significance level $\alpha \in (0,1)$. Since these curves vary in two dimensions, careful consideration is needed to ensure the resulting region represents $\alpha$ in a straightforward and interpretable manner. 

\begin{figure}[!htbp]
    \centering
    \includegraphics[width=.8\textwidth]{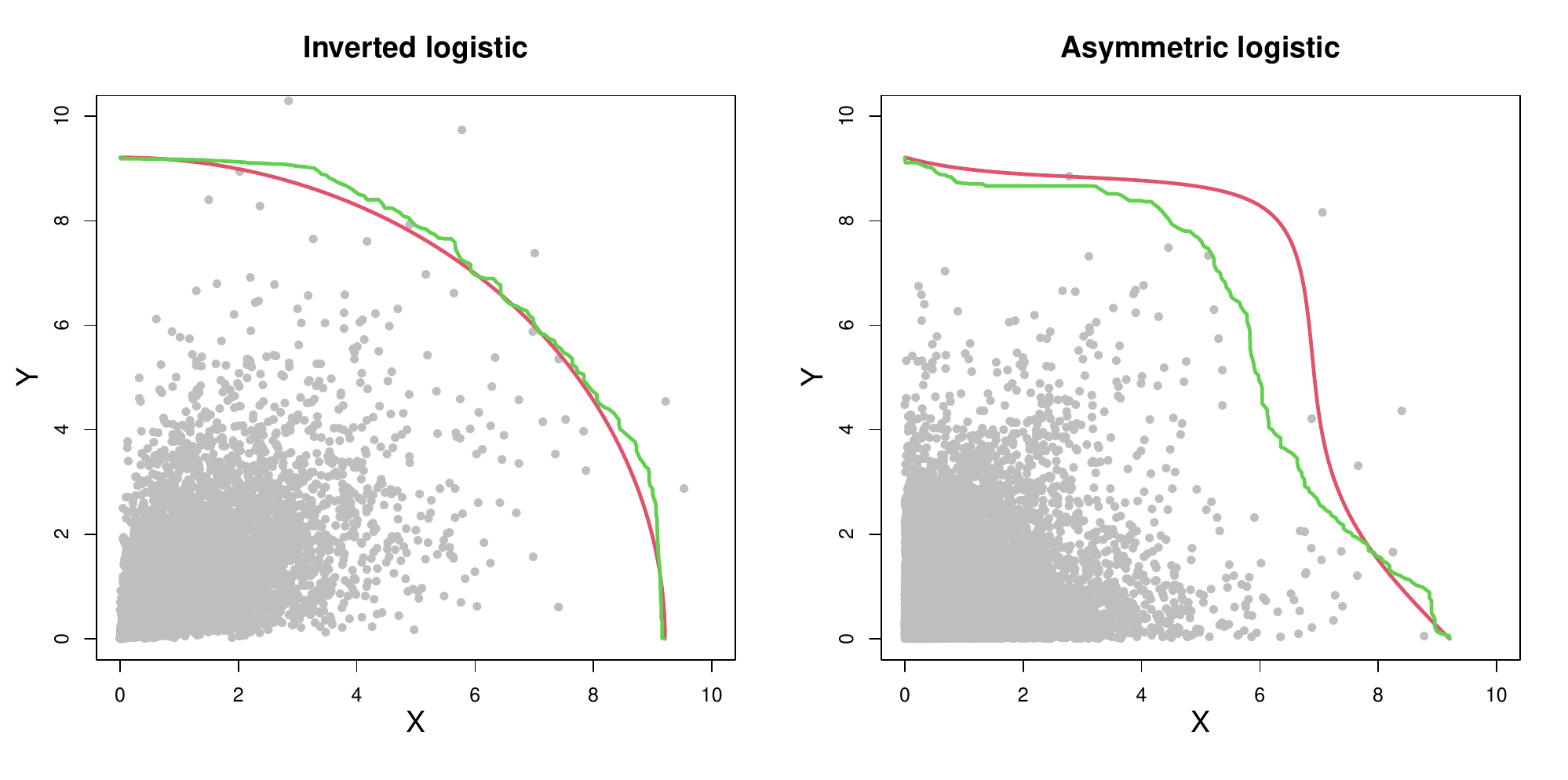}
    \caption{Estimated (green) vs true (red) curves for data sets simulated from inverted logistic (left) and asymmetric logistic (right) copulas.}
    \label{fig:estvstrue}
\end{figure}

Figure \ref{fig:estvstrue} displays $n=10000$ datapoints from inverted logistic \citep{Ledford1997} and asymmetric logistic \citep{Tawn1988} copulas on standard exponential margins. The true return curves for $p=1/10000$ are given in red while the curves estimated using the \citet{Wadsworth2013} model are given in green. A representation of sampling uncertainty will help to determine the quality of these estimates. To achieve this, we propose an adaptation of a tool given in \citet{Haselsteiner2019} for representing uncertainty in environmental contour estimates. The novelty in our approach comes from the fact the original tool has not been directly applied for return curve estimation, even though return curves and environmental contours bear many similarities \citep{Haselsteiner2021}. Moreover, no consideration is given to the theoretical justification of the resulting uncertainty representation in the original approach in terms of coverage properties.

On standard exponential margins, the joint support is given by the set $\RR^2_+$; we consider a set of angles in the interval $(0,\pi/2)$ and use these angles to divide the $\RR^2_+$ plane. Specifically, given some large positive integer $m$, we define $\boldsymbol{\Theta}:= \{\pi (m+1-j)/2(m+1) \mid 1\leq j \leq m\}$, i.e., a sequence of decreasing angles starting near $\pi/2$ and approaching $0$. For each $\theta \in \boldsymbol{\Theta}$, let $L_{\theta} := \{(x,y) \in \RR_+^2 \mid \tan(\theta) = y/x\}$ denote the line segment intersecting the origin with gradient $\tan(\theta)>0$. For any return curve estimate $\widehat{\mathrm{RC}}(p)$ satisfying properties \ref{prop:marg_limit} - \ref{prop:shape_curve}, we have that $L_{\theta}$ intersects $\widehat{\mathrm{RC}}(p)$ exactly once for every $\theta \in \boldsymbol{\Theta}$, implying there is a one-to-one correspondence between angles and points on the estimated curve. An illustrative figure of this correspondence can be found in the supplementary material. 

Letting $\{(\hat{x}_{\theta},\hat{y}_{\theta})\} := \widehat{\mathrm{RC}}(p) \cap L_{\theta}$, we let $\hat{d}_{\theta}$ denote the $l_2$-norm of this point estimate, i.e., $\hat{d}_{\theta}:= (\hat{x}^2_{\theta}+\hat{y}^2_{\theta})^{1/2}$. Since the angle $\theta$ is fixed, this metric represents the aspect of $(\hat{x}_{\theta},\hat{y}_{\theta})$ that will vary across different curve estimates. Uncertainty in return curve estimates can consequently be quantified using the distribution of $\hat{d}_{\theta}$ at each angle $\theta \in \boldsymbol{\Theta}$. We propose the following bootstrap procedure: for $k = 1, \hdots, K$,  
\begin{enumerate}
    \item Bootstrap the original data sample to produce a new sample of the same size. 
    \item For each $\theta \in \boldsymbol{\Theta}$, obtain the $l_2$-norm for the corresponding point estimate obtained using a given model. Denote this value by $\hat{d}_{\theta,k}$. 
\end{enumerate}
If temporal dependence is shown to exist in the data, block bootstrapping can be used for this procedure, allowing one to to account for additional uncertainty that arises. Given $\theta \in \boldsymbol{\Theta}$, we construct empirical estimates of the mean, median, and $100(1-\alpha)\%$ confidence intervals for the $l_2$-norm values using the sample $\{\hat{d}_{\theta,k}\mid 1 \leq k \leq K\}$. Taking $\alpha = 0.95$, we estimate the $2.5\%$ and $97.5\%$ quantiles using this sample, which we denote $\hat{d}_{\theta}^{0.025}$ and $\hat{d}_{\theta}^{0.975}$ respectively. Assuming unbiased estimation, $\Pr (\hat{d}_{\theta}^{0.025} \leq d_{\theta} \leq \hat{d}_{\theta}^{0.975}) \approx 0.95$, where $d_{\theta} = (x_{\theta}^2 + y_{\theta}^2)^{1/2}$ is the $l_2$-norm of $(x_{\theta},y_{\theta}) \in \RC{p}{} \cap L_{\theta}$. Hence, one can show that 
\begin{equation*}
    \Pr\left[  (x_{\theta},y_{\theta}) \in  \left\{ (x,y) \in L_{\theta}  \Big\vert  d_{\theta} \in [\hat{d}_{\theta}^{0.025},\hat{d}_{\theta}^{0.975}] \right\} \right] \approx 0.95, 
\end{equation*}
implying the set $\left\{ (x,y) \in L_{\theta}  \big\vert  d_{\theta} \in [\hat{d}_{\theta}^{0.025},\hat{d}_{\theta}^{0.975}] \right\}$ defines a confidence region for curve points along the line $y = \tan(\theta) x$. Taking the maximum and minimum $x$ and $y$ coordinates in this set, we obtain a pointwise confidence region for points along the line segment $L_{\theta}$ at each angle $\theta \in \boldsymbol{\Theta}$. These pointwise confidence regions, along with the $x$ and $y$ coordinates corresponding to the mean and median $l_2$-norm values, can be joined together in order of angle to construct estimates that represent mean, median, and $95\%$ confidence interval estimates for the return curve. 

\begin{figure}
    \centering
    \includegraphics[width=.8\textwidth]{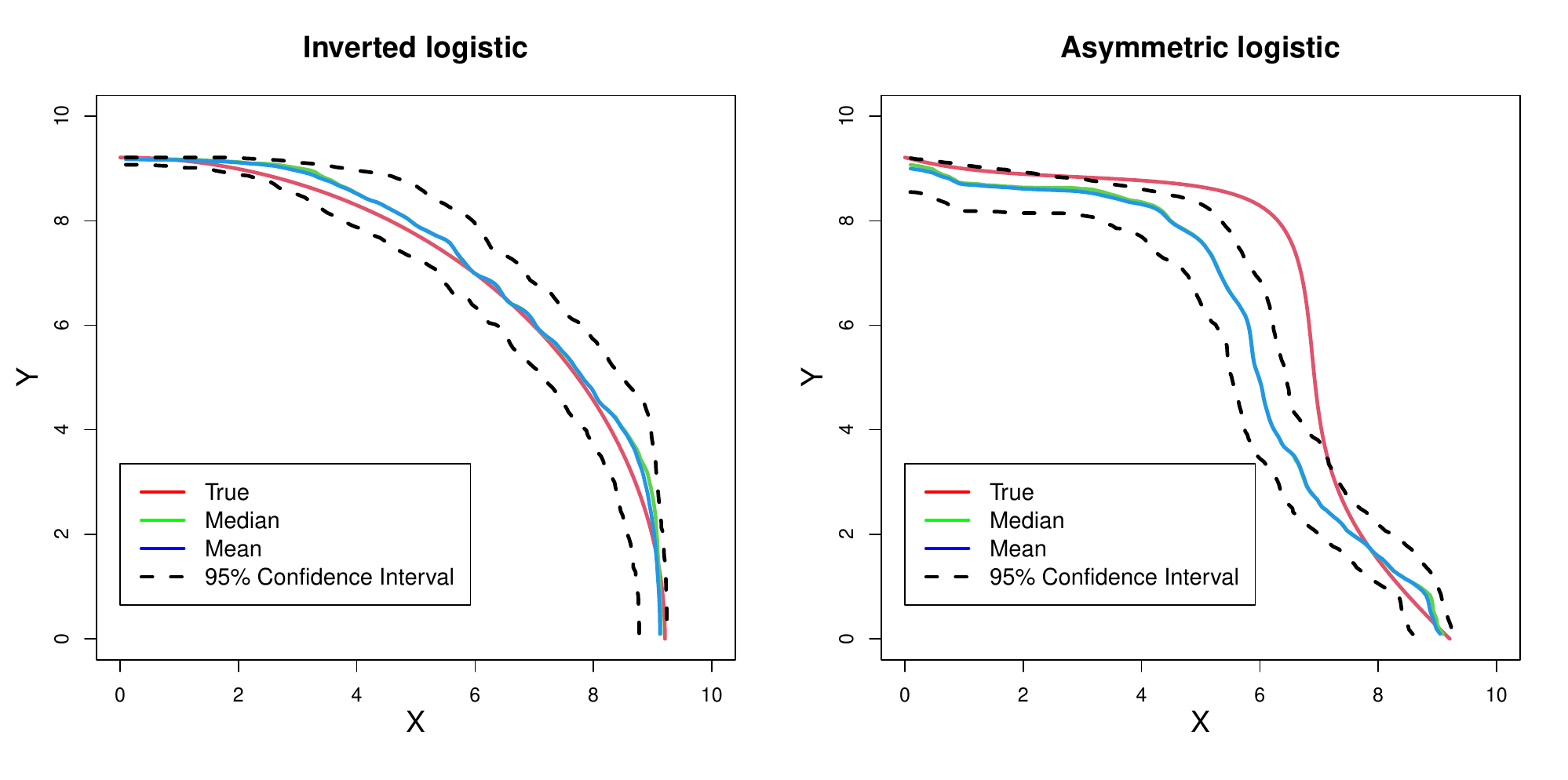}
    \caption{Comparison of median, median, and $95\%$ confidence interval return curve estimates (green, blue and black dotted lines respectively) to the true return curves (red) for inverted logistic (left) and asymmetric logistic (right) copulas with $K=1000$ bootstrap samples.}
    \label{fig:RC_uncertainty}
\end{figure}

Our procedure is illustrated in Figure \ref{fig:RC_uncertainty} with $m=150$. The confidence interval width appears to vary over angles in both cases - this is partly explained by implementation of Properties \ref{prop:marg_limit} - \ref{prop:shape_curve} in each bootstrap curve estimate. For the inverted logistic copula, the true curve is captured by the estimated confidence region at all angles. For the asymmetric logistic copula, the estimated confidence region only captures the true curve in certain regions of the $\RR_+^2$ plane. This observation indicates some bias may exist for curve estimates from this model. This bias is likely a result of the modelling framework being unable to account for the complex asymmetric structure of this copula at finite levels due to a poor rate of convergence to the limiting angular dependence function introduced in Equation \eqref{eqn:wads_tawn}, which is required for return curve estimation. 


\subsection{Return curve diagnostic tool}
\label{subsec:diagnostic}
Since the true return curve is unknown in practice, we require a means of evaluating the goodness of fit for a curve estimate, $\widehat{\mathrm{RC}}(p)$, obtained from a particular sample. We propose such a technique and illustrate the method using a single data set simulated from a logistic copula \citep{Tawn1988} on standard exponential margins. This tool provides a means to assess the accuracy of a given curve estimate for a data set with no knowledge of marginal or copula distributions.

\begin{figure}[!htbp]
    \centering
    \includegraphics[width=.8\textwidth]{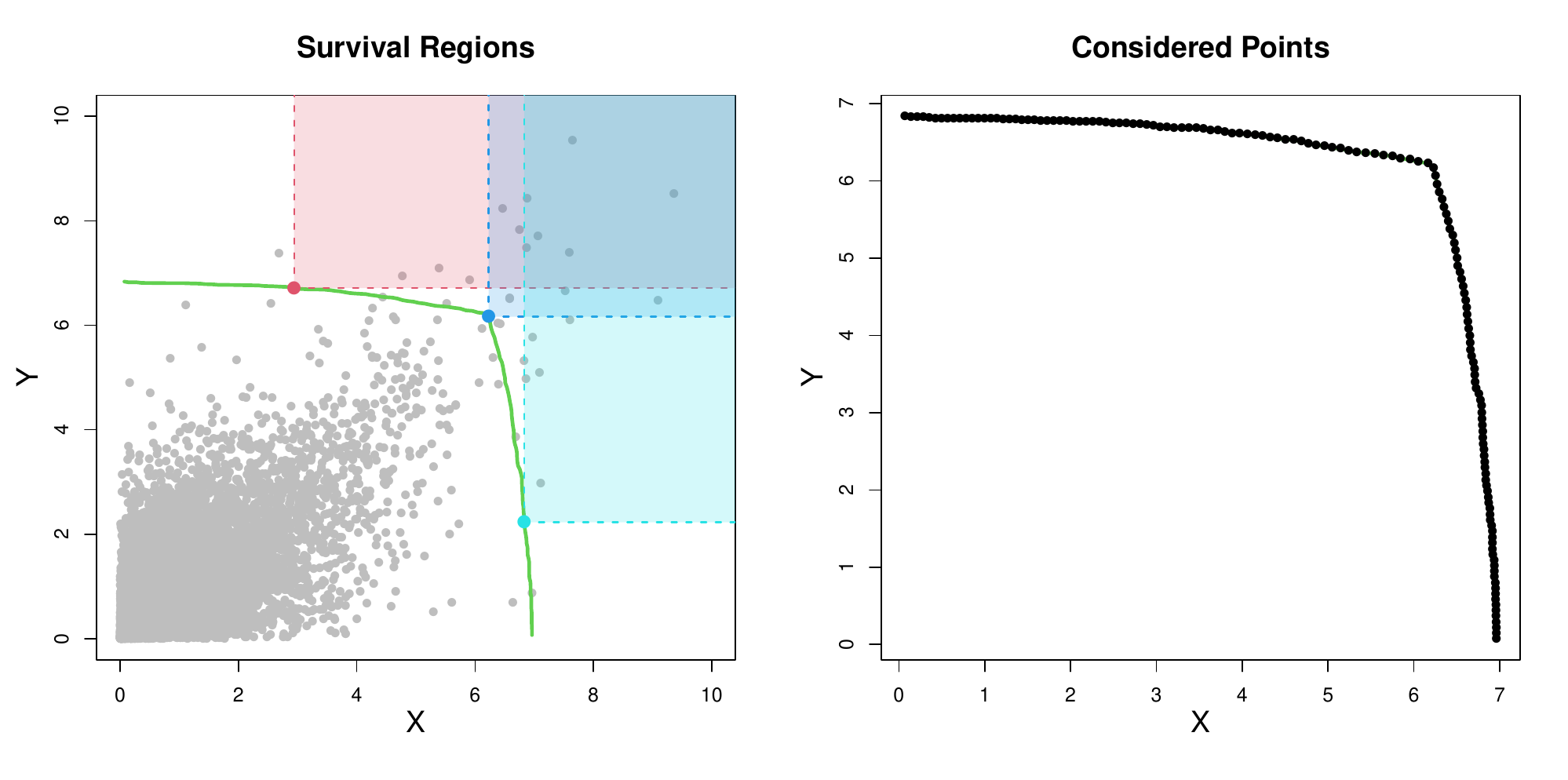}
    \caption{Left: Survival regions for three points on the estimated return curve. Right: Subset of points on the estimated curve considered in diagnostic.}
    \label{fig:surv_region}
\end{figure}

Consider the shaded survival regions defined in the left panel Figure \ref{fig:surv_region} for an estimated return curve $\widehat{\mathrm{RC}}(p)$, where $p$ is small but $\widehat{\mathrm{RC}}(p)$ is in the range of the data. Regions of the form $(x,\infty)\times (y,\infty)$ are illustrated at three different points on the curve. The probability of lying within each such region should, by definition, equal $p$. To assess this, we consider fixed survival regions for a chosen subset of points on the estimated curve. For convenience, this subset is chosen such that points correspond to the set of angles $\boldsymbol{\Theta}$; again we take $m=150$. This results in the set of points sufficing as a representation of the entire estimated curve, as demonstrated in the right panel of Figure \ref{fig:surv_region}. If this estimated curve accurately reflects the true return curve, the empirical probability of observing data within each survival region should be close to $p$. 

Let $j \in [1,150]$ denote the index of an angle $\theta_j \in \boldsymbol{\Theta}$, and let $(\hat{x}_{\theta_j},\hat{y}_{\theta_j})$ denote the corresponding point on the estimated curve. Furthermore, let $(\mathbf{x},\mathbf{y}) = \{(x_i,y_i): 1 \leq i \leq n \}$ denote the observed sample that has been used to estimate the curve. The empirical estimate, which we denote $\hat{p}_j$, is given by the proportion of points lying in the region $(\hat{x}_{\theta_j},\infty)\times (\hat{y}_{\theta_j},\infty)$. We then apply the bootstrap to resample the original data set and this estimation procedure is repeated to obtain a range of empirical estimates. As in Section \ref{subsec:quantifying_uncertainty}, block bootstrapping should be applied if temporal dependence is shown to exist in the data set. For each $j$, we let $\hat{\mathcal{P}}_j$ denote the set of empirical probability estimates obtained using bootstrapping. Finally, we estimate the median and $95\%$ pointwise confidence intervals for the probabilities at index $j$ by taking empirical $2.5\%$, $50\%$ and $97.5\%$ quantiles of the set $\hat{\mathcal{P}}_j$. These estimates provide a pointwise diagnostic at each angle, and can be combined over angles to represent the diagnostic procedure over the whole curve. 

\begin{figure}[!htbp]
    \centering
    \includegraphics[width=.4\textwidth]{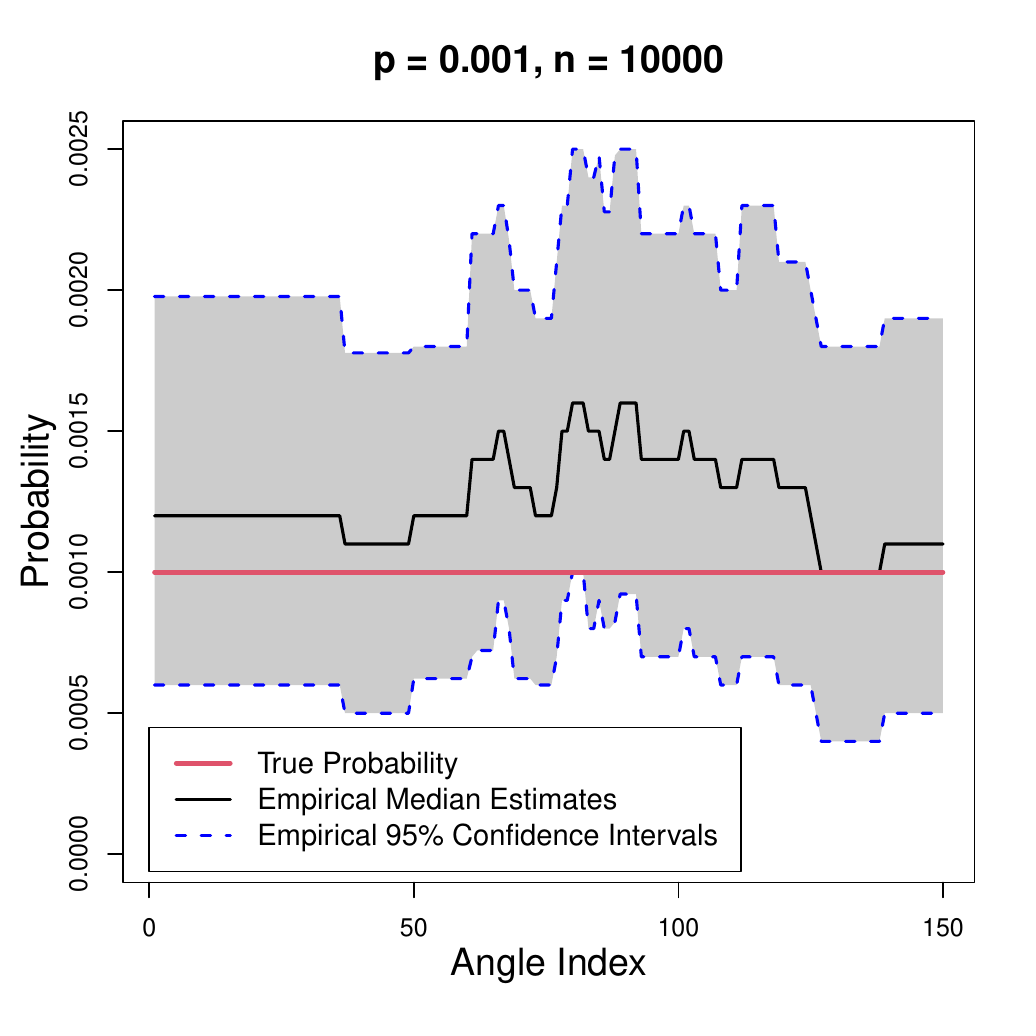}
    \caption{Illustration of diagnostic tool. Solid red and black lines denotes true and mean empirical estimates, respectively, and grey shaded region between dotted blues lines describe empirical 95\% CI estimates.}
    \label{fig:diag_example}
\end{figure}

This procedure is illustrated in Figure \ref{fig:diag_example} using the example given in Figure \ref{fig:surv_region}. The black line and shaded regions in the figure represent the empirical estimates of the median and $95\%$ pointwise confidence intervals, respectively, for each index, with the red line denoting the true probability. As can be observed, for all indices, the confidence bounds contain the true value $p$, suggesting this estimated curve accurately represents this value. However, the median empirical estimates are greater than $p$ at the majority of indices, suggesting a slight overestimation bias for this particular curve estimate. 

We note that the confidence intervals produced through bootstrapping for the return curves in Section \ref{subsec:quantifying_uncertainty}, and survival probabilities in Section \ref{subsec:diagnostic}, are all pointwise and dependent across angles. Although they cannot be interpreted across the whole range, they still provide a useful assessment of the utility of various curve estimation techniques. Both tools are adapted in Section \ref{Sec6} to account for the original margins of the environmental data sets, allowing us to analyse the quality of return curve estimates for these examples.

\section{Simulation study}\label{Sec5}

We compare the return curve estimates from the models discussed in Section \ref{subsec:novelRCestimation} to those estimated using the methodology of \citet{Cooley2019}. For this, we consider several simulated data sets on standard exponential margins, representing a range of different extremal dependence structures. Specifically, we consider the following copula families: logistic and asymmetric logistic copulas from the bivariate extreme value (BEV) family, the bivariate normal copula with correlation coefficient $\rho$, logistic and asymmetric logistic copulas from the inverted BEV family, the bivariate t copula with correlation coefficient $\rho$ and degrees of freedom $\nu$ and the Frank copula with dependence parameter $\zeta$.

For the methods introduced in \citet{Cooley2019}, we transform data to standard Fr\'echet margins, use the procedures proposed in the paper to obtain return curve estimates, transform back to standard exponential margins and apply Properties \ref{prop:marg_limit}, \ref{prop:curve_limits} and \ref{prop:shape_curve}. For each example, the chosen estimation procedure is determined by the extremal dependence exhibited by the underlying copula; this must be specified prior to inference, illustrating a drawback of this approach. We use our knowledge of the true dependence structure to implement the correct procedure, but in practice such knowledge would not be available to us. The code for implementing this approach can be found at \url{https://www.stat.colostate.edu/~cooleyd/Isolines/}. 

\begin{figure}[!ht]
\centering
\includegraphics[width=.9\textwidth]{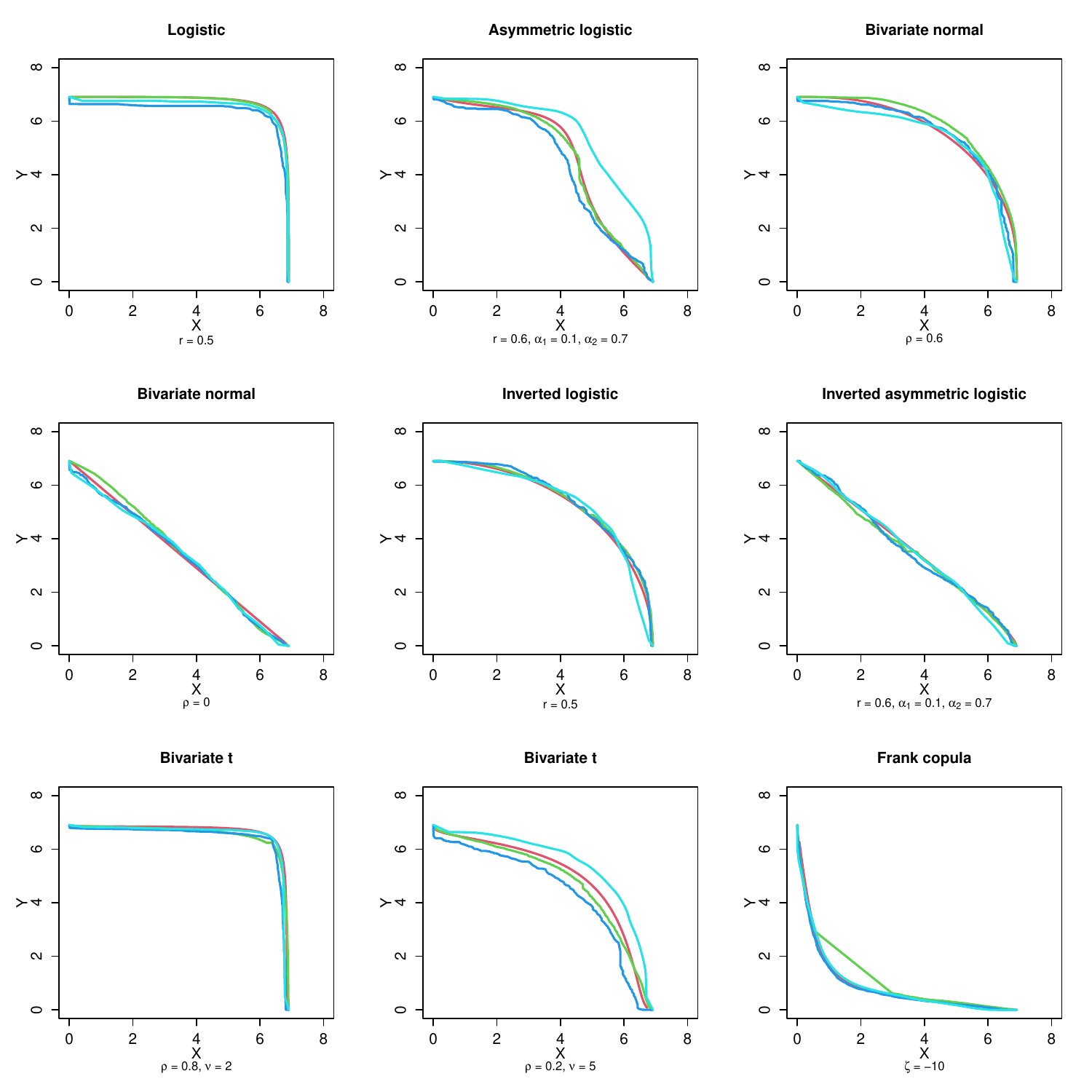}
\caption{Examples curve estimates for each copula family. True curves are given in red, while the estimated curves from the \citet{Heffernan2004}, \citet{Wadsworth2013} and \citet{Cooley2019} models are given in green, dark blue and light blue, respectively.}
\label{fig:copula_examples}
\end{figure}

Examples of both estimated and true return curves for each copula, with $n=10000$ and $p=10^{-3}$, are illustrated in Figure \ref{fig:copula_examples}. We note that for the Frank copula, there is a distinct `linear' segment of the curve estimate from the \citet{Heffernan2004} model: this lack of fit reflects a shortcoming of this approach for data sets with negative dependence. Since the modelling framework requires us to condition on either $X_L$ or $Y_L$, we can only evaluate joint tail behaviour in the region where at least one variable is large, i.e., $\{ (x,y) \in \RR^2 \mid x>u_{X_L} \; \text{OR} \; y>u_{Y_L}\}$. As observed for the Frank copula, part of the true return curve can be defined outside of this region for negatively dependent data sets; consequently, this curve region cannot be estimated using the \citet{Heffernan2004} framework. This explains the linear segment, since the point estimates for the regions $R_{Y_L>X_L}$ and $R_{Y_L\leq X_L}$ are connected to obtain $\widehat{\mathrm{RC}}(p)$.

To compare curve estimates, we evaluated bias, computational cost and sampling uncertainty for each of the three curve estimation procedures. To assess bias, $1000$ samples of size $n=100000$ were simulated from each copula and the probabilities $p=10^{-3}$ and $p=10^{-4}$ were considered. Adapting the procedure detailed in Section \ref{subsec:quantifying_uncertainty}, median curve estimates for each copula were obtained over the $1000$ samples and compared to the corresponding true curves. While a median curve cannot be computed in practice, this measure should provide a summary of the bias that arises from each procedure. 


To summarise results, we consider a plot of the $l_2$-norm values at angles $\theta_j \in \boldsymbol{\Theta}$ against the corresponding indices $j \in [1,m]$ for the true and estimated median curves, with $m=150$. Numerical methods can then be used to compute the absolute area between the resulting norm curves, with smaller area values corresponding to median curves with less bias. An illustration of this procedure can be found in the supplementary material.
 


The summary statistics of integrated absolute difference for each copula-model pairing are given in Table \ref{table:summary_stat}. The bias from each procedure appears to vary significantly over the different copula structures, suggesting that the bias in return curve estimates exhibited by a particular model varies with the form of extremal dependence. It is clear that the bias from the \citet{Cooley2019} curve estimates are significantly larger for all but three of the copulas considered; however, we note than in all cases, this method has an unrealistic advantage, namely that the extremal dependence classes have been correctly specified. On the other hand, the models proposed in \citet{Heffernan2004} and \citet{Wadsworth2013} appear to have similar amounts of bias across the majority of copula structures considered, and neither consistently outperforms the other. Since models for multivariate extremes are typically based on asymptotic arguments which sometimes hold better for one data set than another, this conclusion is most likely a reflection of the different asymptotic arguments for these models.

To evaluate the computational cost of each estimation technique, fifty samples of size $n=10000$ were simulated from a logistic copula with dependence parameter $0.5$. With $p=10^{-3}$, a Windows machine with a 1.60 GHz Intel(R) Core(TM) i5-8250U processor and 16GB of RAM was used to compute return curve estimates for each of the fifty samples, and the total computation times were recorded. For the \citet{Heffernan2004}, \citet{Wadsworth2013} and \citet{Cooley2019} techniques, these times were 269.9s, 6.1s, and 2618.2s, respectively.

Application of the \citet{Wadsworth2013} modelling framework was significantly quicker than the other two approaches; this is likely due to the fact this technique does not involve any simulation and/or smoothing. Of the remaining estimation frameworks, application of the \citet{Heffernan2004} model was still significantly quicker than the method given in \citet{Cooley2019}. This conclusion appears to be a result of the Gaussian-kernel density smoothing techniques that are applied when obtaining the empirical curve estimates for the latter approach. Combined with the fact the bias appears significantly lower for the other two models, we choose not consider the approach of \citet{Cooley2019} further.

To assess the sampling uncertainty from the remaining procedures, we compute the coverage for estimated confidence regions at fixed angles. For this, $500$ simulated samples of size $n=10000$ from each copula were considered. Using bootstrapping with $K=200$ iterations, confidence regions were obtained following the procedure outlined in Section \ref{subsec:quantifying_uncertainty} for probabilities of $p=10^{-3}$ and $10^{-4}$, and we assessed the coverage of these at five fixed angles $\theta \in \{\pi (m+1-j)/2(m+1) \mid j = 1, 38, 75,112,150  \}$, allowing assessment of coverage for a variety of regions. Two of the angles are only considered for the BEV asymmetric logistic and inverted BEV asymmetric logistic copulas, since these are the only distributions not to exhibit symmetry. We consider $95\%$ confidence regions for both probabilities. The results for $p=10^{-3}$ are given in Table \ref{table:cov1}; the results for $p=10^{-4}$ can be found in the supplementary material, along with an visual illustration of the coverage procedure.

These coverage results provide an insight into differences in the \citet{Heffernan2004} and \citet{Wadsworth2013} models. Firstly, for angles close to $0$ and $\pi/2$, the coverage from the \citet{Wadsworth2013} model tends to be closer to the nominal level then that from the \citet{Heffernan2004} model. This is especially apparent when examining the scores at both probabilities for the logistic and first bivariate normal copula examples. We note that imposing Property \ref{prop:marg_limit} will affect the coverage near the margins, since we do not allow return curve coordinate estimates that exceed the marginal $(1-p)$-th quantiles, resulting in constrained confidence intervals. We also note that the coverage values for the Frank copula from the \citet{Heffernan2004} framework are noticeably small; this relates to aforementioned shortcoming of this approach for data sets with negative dependence. On the other hand, certain coverage values obtained using the \citet{Wadsworth2013} approach are noticeably smaller than the corresponding values from the \citet{Heffernan2004} approach; for example, for the BEV asymmetric logistic copula at $p=10^{-4}$ and the second bivariate t copula. On the whole, neither procedure consistently outperforms the other over the copulas and angles considered and encouragingly, the resulting coverage scores were, in many cases, close to the nominal level.

From these results, we suggest that the curve estimation technique derived using the \citet{Wadsworth2013} model is preferable in a practical setting; it is straightforward to implement and significantly outperforms the other considered techniques in terms of computation time. Combined with bias and coverage results, alongside the ability of the \citet{Wadsworth2013} model to capture negative dependence structures, this curve estimation technique offers clear advantages over the alternative methods, making it the best suited for practical applications.

\section{Case study}\label{Sec6}

\begin{figure}[!ht]
    \centering
    \includegraphics[width=.8\textwidth]{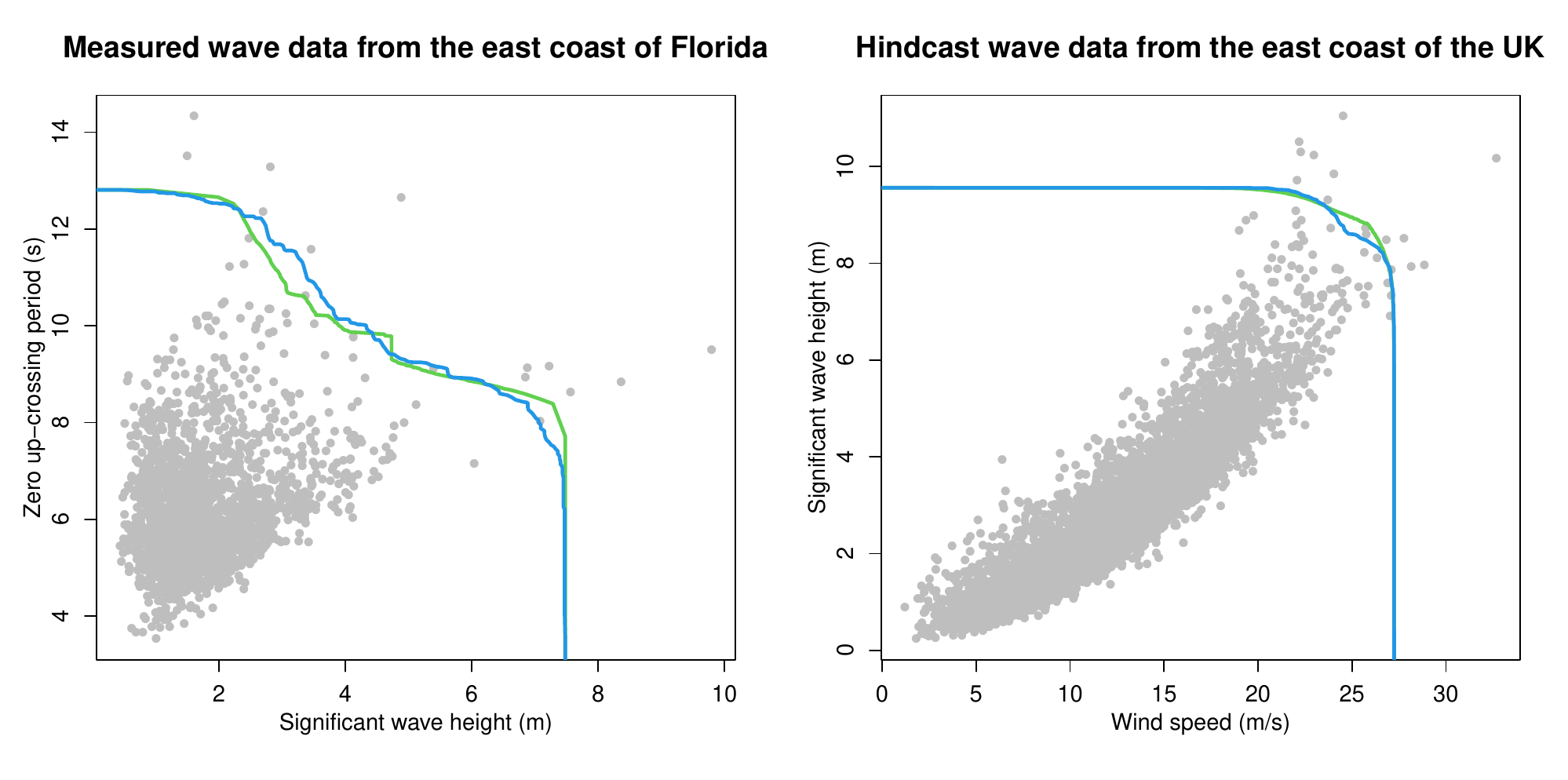}
    \caption{Return curves estimated for measured (left) and hindcast (right) data sets, with $p=10^{-3}$. Green and blue lines represent the estimates from the \citet{Heffernan2004} and \citet{Wadsworth2013} models, respectively.}
    \label{fig:case_return_curves}
\end{figure}

We now apply our techniques to the two metocean data sets introduced in Section \ref{Sec1}. We first transform both data sets to standard exponential margins. Assuming each margin is identically distributed over time, we estimate the marginal distributions using equation \eqref{eqn:CDF_trans} and apply the probability integral transform. We then use the techniques proposed in Section \ref{subsec:novelRCestimation} to obtain curve estimates for the probability $p=10^{-3}$, corresponding to a return period of approximately $4.7$ years. The resulting curve estimates are illustrated in Figure \ref{fig:case_return_curves} on the original margins (following back transformation). 

Next, we adapt the diagnostic tool introduced in Section \ref{Sec4} to assess the goodness of fit of return curve estimates for both data sets. To account for the additional uncertainty that arises during estimation of the marginal distributions, we apply the diagnostic on the original margins of the data. This is done as follows: letting $\{(x_i,y_i)\}_{i=1,\ldots,n}$ denote either data set, we define $(x_0,y_0) = (\min_ix_i,\min_iy_i)$. We use these coordinates as a reference point from which we can evaluate return curve estimates. Given $\boldsymbol{\Theta}$ defined as before, we define the line segment $L'_{\theta} := \{(x,y) \in \RR^2 \mid y = (x - x_0)\tan(\theta) + y_0\}$ for each $\theta \in \boldsymbol{\Theta}$ and, for any curve estimate $\widehat{\mathrm{RC}}(p)$, consider the intersection of the sets $\widehat{\mathrm{RC}}(p) \cap L'_{\theta}$. Illustrations of reference points, line segments and intersection points are given in the supplementary material for both data sets. Similarly to before, these intersection points are used to define the joint survival regions; the data is then resampled and sets of empirical probability estimates are obtained for each region. 

\begin{figure}[!ht]
    \centering
    \includegraphics[width=.8\textwidth]{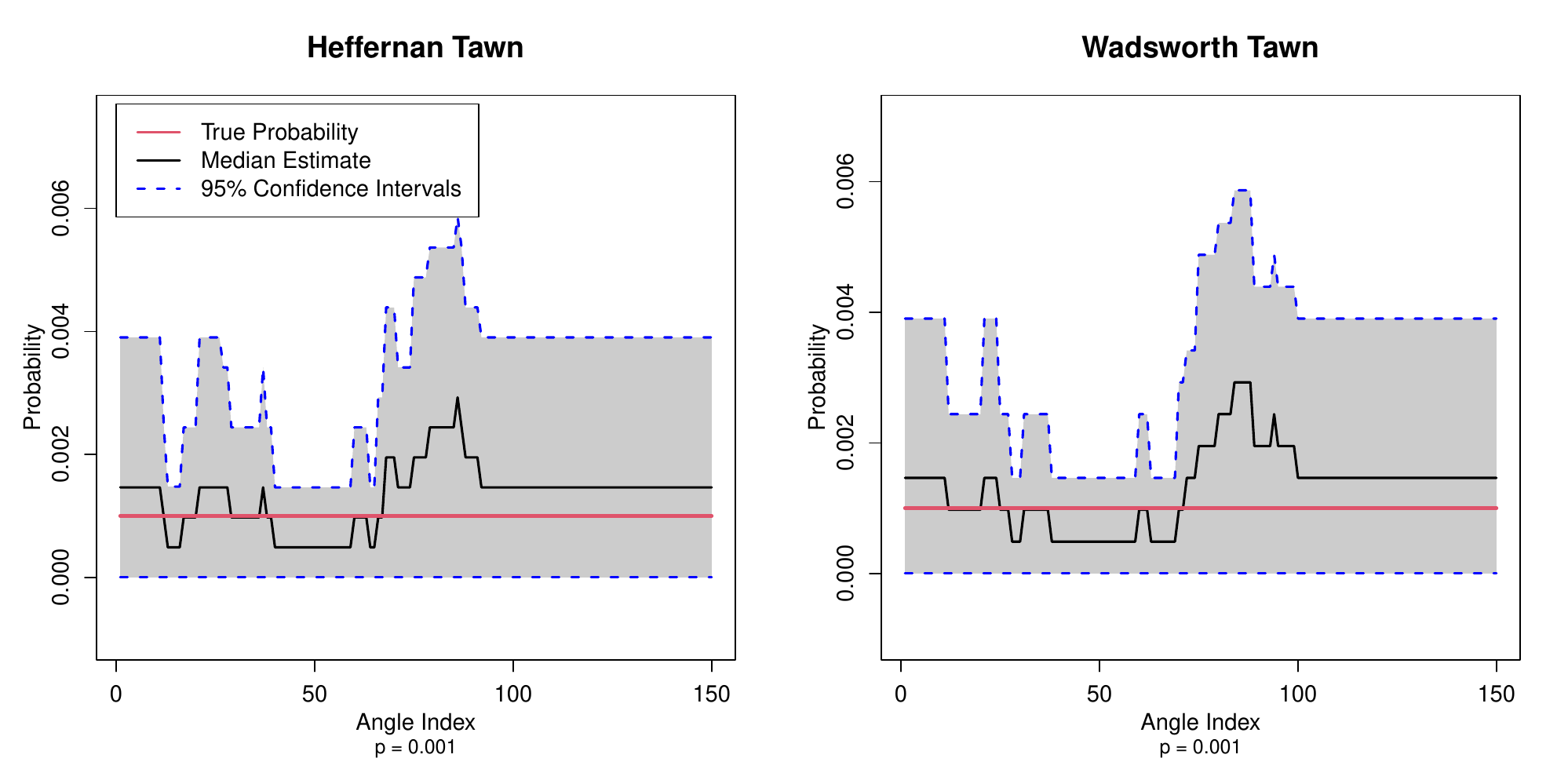}
    \caption{Measured data diagnostic plots with $K=1000$ block bootstraps from \citet{Heffernan2004} (left) and \citet{Wadsworth2013} (right) models, respectively.}
    \label{fig:casediag1}
\end{figure}

\begin{figure}[!ht]
    \centering
    \includegraphics[width=.8\textwidth]{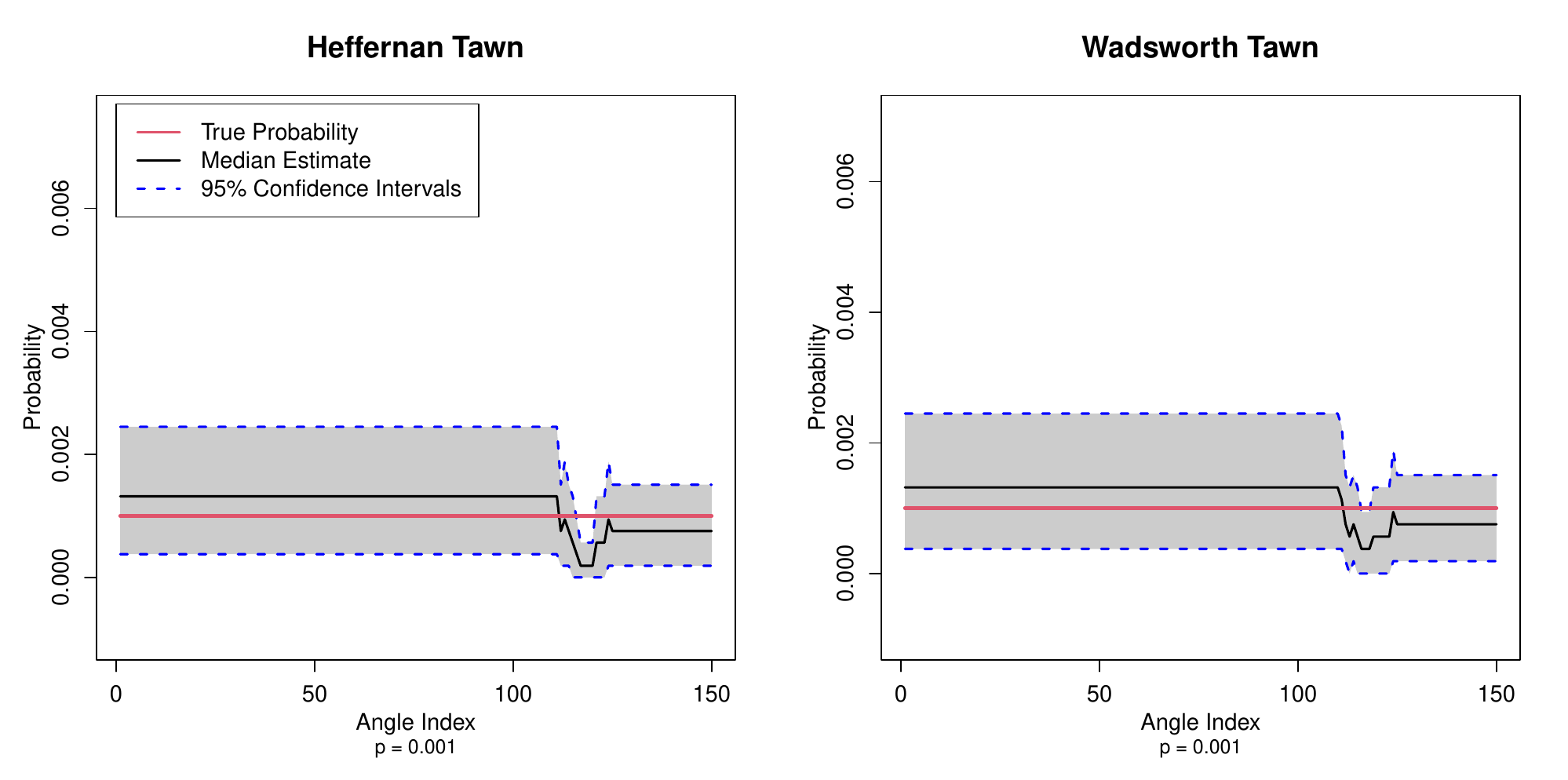}
    \caption{Hindcast data diagnostic plots with $K=1000$ block bootstraps from \citet{Heffernan2004} (left) and \citet{Wadsworth2013} (right) models, respectively.}
    \label{fig:casediag2}
\end{figure}

We apply block bootstrapping for resampling, because even with pre-processing, both data sets still appear to exhibit some marginal temporal dependence. Block sizes of 5 and 10 were selected for the measured and hindcast data sets, respectively, by considering plots of the autocorrelation function and selecting values beyond which the dependence appeared insignificant for both variables. These block sizes were then used to bootstrap the original data sets. The resulting diagnostic plots are given in Figures \ref{fig:casediag1} and \ref{fig:casediag2}. 


For the observed data, Figure \ref{fig:casediag1} suggests both models perform similarly and provide accurate curve estimates, while for the hindcast data, Figure \ref{fig:casediag2} suggests the \citet{Wadsworth2013} curve estimate outperforms the \citet{Heffernan2004} estimate at a subset of angles. This is reflected by the difference in curve estimates in the the joint upper tail of the data; see Figure \ref{fig:case_return_curves}. In both cases, the variability in curve estimates appears to vary with the angle. The estimated confidence intervals capture the true probability at the majority of considered angles, suggesting both estimation techniques can accurately capture joint tail behaviour for these data sets.

\begin{figure}[!ht]
    \centering
    \includegraphics[width=.8\textwidth]{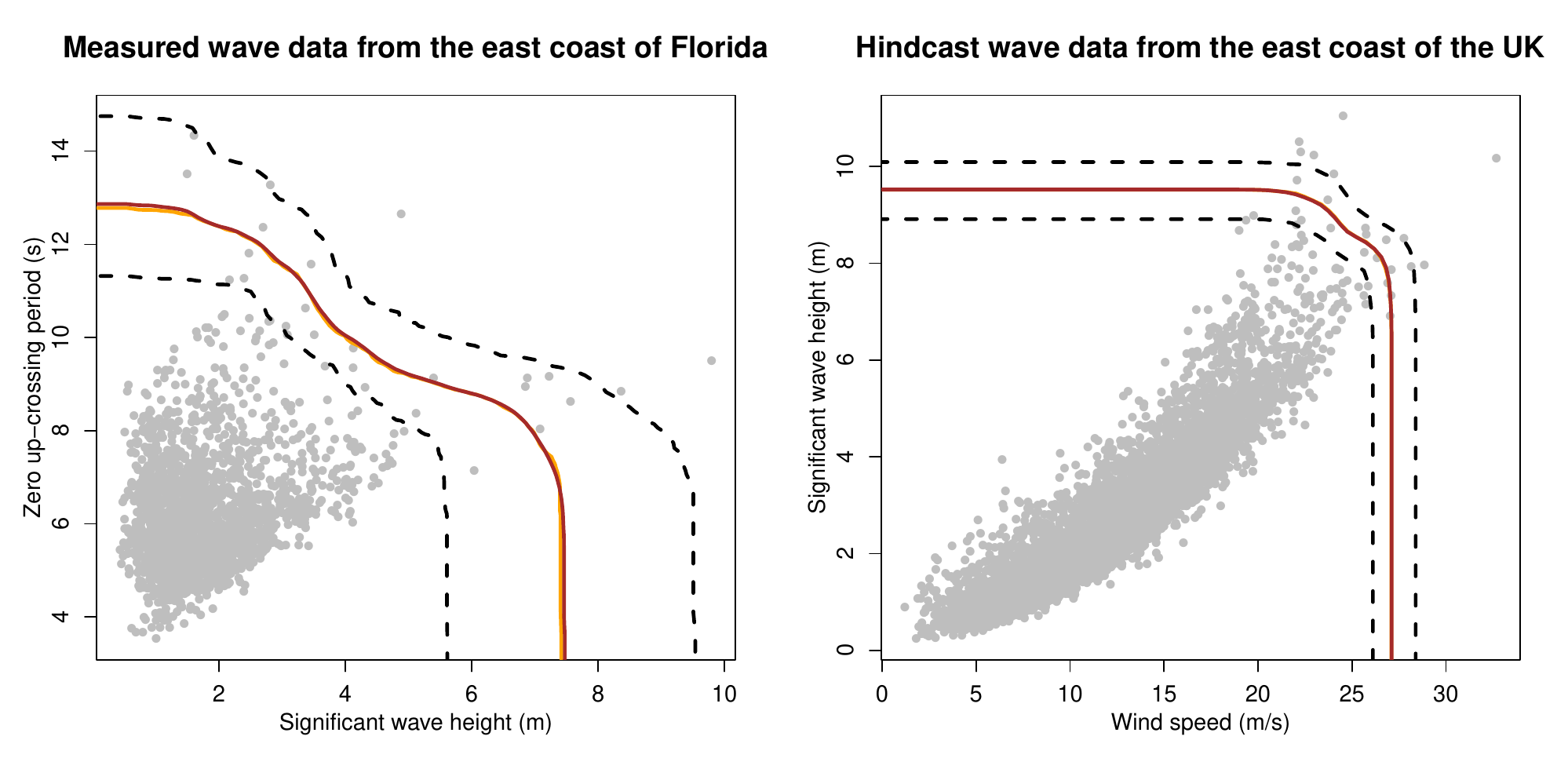}
    \caption{Median (orange) and mean (brown) curve estimate, along with 95\% (black dotted) confidence regions obtained using block bootstrapping with $K=250$ for the measured (left) and hindcast (right) data sets. The \citet{Wadsworth2013} model was used to obtain all estimates.}
    \label{fig:case_uncert}
\end{figure}

Finally, we apply an adaptation of the technique introduced in Section \ref{subsec:quantifying_uncertainty} to assess uncertainty in return curve estimates. Letting $(\hat{x}_{\theta},\hat{y}_{\theta}) := \widehat{\mathrm{RC}}(p) \cap L'_{\theta}$, with $L'_{\theta}$ defined as before, uncertainty is captured by considering the distribution of $l_2$-norm values from the reference point, i.e., $\hat{d}_{\theta} = \lvert (\hat{x}_{\theta}-x_0)^2 + (\hat{y}_{\theta}-y_0)^2 \rvert^{1/2}$. Using block bootstrapping, with the marginal distributions re-estimated for each bootstrapped sample, we obtain median and mean curve estimates, along with pointwise $95\%$ confidence intervals across angles. For the reasons outlined in Section \ref{subsec:diagnostic}, we use the \citet{Wadsworth2013} approach to obtain these estimates. The resulting curve estimates are illustrated in Figure \ref{fig:case_uncert}; we observe that taking mean and median curves appears to have a smoothing effect on the resulting estimates, relative to the original curve estimates given in Figure \ref{fig:case_return_curves}.

\section{Discussion}\label{Sec7}
We have considered the concept of a return curve as a bivariate extension to a return level and introduced novel estimation techniques, illustrating that these methods perform better than an existing approach. Furthermore, unlike \citet{Cooley2019}, our methods do not require the form of extremal dependence to be pre-specified: this is an obvious advantage, since determining the extremal dependence structure is seldom straightforward. We have also proposed novel uncertainty representation and diagnostic tools for the risk measure.

For the diagnostic tool proposed in Section \ref{subsec:diagnostic}, we note that extreme survival region probabilities are estimated empirically, meaning the accuracy of such estimates will be directly related to the sample size $n$ and the probability $p$. This represents a broader problem within the extremes literature, since, by definition, we will have observed very few extremes values that can be used to verify and justify a given approach. This issue is illustrated further in the supplementary material, where we consider decreasing probabilities for a fixed sample size. However, as with similar analyses, if the tool appears to illustrate a good fit to the data at less extreme probabilities, we can be more confident in extrapolating to more extreme values. 

While we have focused on bivariate random vectors, both concepts and methodology can be extended to the general multivariate setting. However, higher dimensional curves are difficult to visualise and capturing the dependence structures in higher dimensions becomes increasingly complex, since different two dimensional marginals can exhibit different forms of extremal dependence. Nevertheless, in the bivariate setting, we believe return curves are a useful tool for researchers to explore joint extremal behaviour and develop a better understanding of potential risks. Indeed, return curves are already utilised to analyse risks for ocean and coastal structures. It is also important to note this risk measure only denotes the rarity of events, not impact: therefore, in practice, researchers must carefully consider which regions of the multivariate space are impactful prior to inference.

As is common in many environmental contexts, the data sets considered in Section \ref{Sec6} both appear to exhibit non-negligible temporal dependence. We account for this feature by using block bootstrapping to quantify uncertainty, but this creates the additional challenge of block size selection. For this, we use an ad-hoc technique based on examinations of ACF plots. An in-depth investigation could improve on this approach through a more robust, theoretically justified resampling scheme. 

For the hindcast data set discussed in Section \ref{Sec6}, Figure \ref{fig:casediag2} illustrates a downside of applying the diagnostic tools on the original marginal distributions. In particular, the majority of angles selected to represent the curve estimates correspond to the `marginal limits' of the curves, i.e., the straight line segments connecting the curve estimates to the margins in Figure \ref{fig:case_uncert}. Due to the strong positive dependence between the hindcast variables, empirical survival probabilities will be unchanging along these line segments, explaining the largely constant diagnostic probabilities in Figure \ref{fig:casediag2}. This is further demonstrated by the illustration of line segments given in the supplementary material. Future research could explore techniques for selecting angles such that the corresponding return curve representation is exclusively in-between the aforementioned marginal limits. 


Finally, we note that all techniques discussed in this paper are only applicable to data sets exhibiting stationarity; accounting for non-stationarity in the context of return curves presents many challenges, since return curves are defined in the stationary setting only and all models introduced in Section \ref{Sec3} assume stationarity. While a range of approaches exist for capturing non-stationary in the univariate setting \citep[e.g.][]{Eastoe2019}, relatively few approaches exist in the multivariate setting. This topic has recently been explored in \citet{Murphy-Barltrop2022}, whereby the authors extend the definition of return curves to the non-stationary setting and provide techniques for their estimation. 

\section*{Acknowledgements}
This paper is based on work completed while Callum Murphy-Barltrop was part of the EPSRC funded STOR-i centre for doctoral training (EP/L015692/1). We are grateful to the two referees for constructive comments that have improved this article.
\bigskip
\begin{center}
{\large\bf SUPPLEMENTARY MATERIAL}
\end{center}
\begin{description}
\item[Supplementary material for ``New estimation methods for extremal bivariate return curves"] 
File containing figures and tables that further illustrate ideas and results discussed in the article. (.pdf file)
\item[Code and data.] Zip file containing two R scripts and the case study data sets. The first script reproduces examples of results from Section \ref{Sec5}, while the second script can be used to reproduce case study results from Section \ref{Sec6}. (.zip file)
\end{description}

\bibliography{ms.bib}

\begin{table}[ht!]
\centering
\caption{Summary statistics for each model under different copula structures. In each case, $1000$ samples of $n=100000$ datapoints were simulated and the median curves were computed for $p=10^{-3}$ and $10^{-4}$. `HT', `WT' and `CO' correspond to the median curve estimates from the \citet{Heffernan2004}, \citet{Wadsworth2013} and \citet{Cooley2019} models, respectively.}
\begin{tabular}{|c|c|c|c|c|c|c|c|}
\hline
\multirow{2}{*}{Copula}          &                & \multicolumn{3}{c|}{$p = 10^{-3}$} & \multicolumn{3}{c|}{$p = 10^{-4}$} \\ \cline{2-8} 
                                 & Model             & HT          & WT         & CO  & HT          & WT         & CO  \\ \hline
BEV Logistic                     & \multirow{9}{*}{} & 1.80    & 10.83  & 0.21     & 2.56    & 14.07  & 0.22    \\ \cline{1-1} \cline{3-8} 
BEV Asymmetric Logistic          &                   & 23.61    & 28.27   & 85.50      & 47.66    & 78.16   & 126.25    \\ \cline{1-1} \cline{3-8} 
Bivariate Normal 1               &                   & 2.49    & 3.01   & 25.31    & 4.07    & 7.69   & 37.21     \\ \cline{1-1} \cline{3-8} 
Bivariate Normal 2               &                   & 0.18   & 0.14  &  8.83    & 0.37   & 0.22  & 13.14   \\ \cline{1-1} \cline{3-8} 
Inverted BEV Logistic            &                   & 2.75   & 0.39  & 26.23     & 4.06    & 0.56   & 34.12    \\ \cline{1-1} \cline{3-8} 
Inverted BEV Asymmetric Logistic &                   & 0.80   & 0.19  &  11.47     & 1.22   & 0.30  &11.35    \\ \cline{1-1} \cline{3-8} 
Bivariate T 1                    &                   & 8.26    & 7.17   & 2.36     & 11.24    & 10.44   & 3.87    \\ \cline{1-1} \cline{3-8} 
Bivariate T 2                    &                   & 14.62    & 26.37   & 65.05    & 39.77    & 74.95   & 90.72    \\ \cline{1-1} \cline{3-8} 
Frank                            &                   & 46.39    & 9.65  & 5.65    & 30.33     & 46.56     & 29.70     \\ \hline
\end{tabular}

\label{table:summary_stat}
\end{table}

\begin{landscape}

\begin{singlespace}
\begin{table}[ht!]
\centering
\caption{Coverage values of $95\%$ confidence regions for $p=10^{-3}$. `HT' and `WT' correspond to the \citet{Heffernan2004} and \citet{Wadsworth2013} models, respectively.}
\begin{tabular}{|c|c|c|c|c|c|c|c|c|c|c|c|}
\hline
\multirow{2}{*}{Copula}                           & Probability & \multicolumn{10}{c|}{$p = 10^{-3}$}               \\ \cline{2-12} 
                                                  & Model       & \multicolumn{5}{c|}{HT} & \multicolumn{5}{c|}{WT} \\ \hline
\multirow{2}{*}{BEV Logistic}                     & Angle    & 1   & 2  & 3  & 4  & 5  & 1   & 2  & 3  & 4  & 5  \\ \cline{2-12} 
                                                  & Coverage    & 0.014    & 0.970    & 0.800   & *  & *  & 0.856     & 0.606   &    0.896 & *  & *  \\ \hline
\multirow{2}{*}{BEV Asymmetric Logistic}          & Angle    & 1   & 2  & 3  & 4  & 5  & 1   & 2  & 3  & 4  & 5  \\ \cline{2-12} 
                                                  & Coverage    & 0.756     & 0.938   & 0.868    & 0.860   & 0.896   & 0.908    & 0.054   & 0.362   & 0.902   & 0.936   \\ \hline
\multirow{2}{*}{Bivariate Normal 1}               & Angle    & 1   & 2  & 3  & 4  & 5  & 1   & 2  & 3  & 4  & 5  \\ \cline{2-12} 
                                                  & Coverage    & 0.358     & 0.942   & 0.956   & *  & *  & 0.872    & 0.930   & 0.912   & *  & *  \\ \hline
\multirow{2}{*}{Bivariate Normal 2}               & Angle    & 1   & 2  & 3  & 4  & 5  & 1   & 2  & 3  & 4  & 5  \\ \cline{2-12} 
                                                  & Coverage    & 0.948    & 0.930    & 0.955   & *  & *  & 0.930    & 0.934    & 0.934   & *  & *  \\ \hline
\multirow{2}{*}{Inverted BEV Logistic}            & Angle    & 1   & 2  & 3  & 4  & 5  & 1   & 2  & 3  & 4  & 5  \\ \cline{2-12} 
                                                  & Coverage    & 0.780     & 0.932   & 0.932   & *  & *  & 0.864     & 0.944   &    0.942 & *  & *  \\ \hline
\multirow{2}{*}{Inverted BEV Asymmetric Logistic} & Angle    & 1   & 2  & 3  & 4  & 5  & 1   & 2  & 3  & 4  & 5  \\ \cline{2-12} 
                                                  & Coverage    & 0.960     &0.929    &0.942    &0.951    & 0.922   & 0.924    &    0.934& 0.958    & 0.942   & 0.904   \\ \hline
\multirow{2}{*}{Bivariate T 1}                    & Angle    & 1   & 2  & 3  & 4  & 5  & 1   & 2  & 3  & 4  & 5  \\ \cline{2-12} 
                                                  & Coverage    & 0.504     & 0.928   & 0.512   & *  & *  & 0.920    & 0.758    & 0.884   & *  & *  \\ \hline
\multirow{2}{*}{Bivariate T 2}                    & Angle    & 1   & 2  & 3  & 4  & 5  & 1   & 2  & 3  & 4  & 5  \\ \cline{2-12} 
                                                  & Coverage    & 0.784     & 0.896   & 0.930   & *  & *  & 0.938     & 0.806    & 0.632   & *  & *  \\ \hline
\multirow{2}{*}{Frank}                            & Angle    & 1   & 2  & 3  & 4  & 5  & 1   & 2  & 3  & 4  & 5  \\ \cline{2-12} 
                                                  & Coverage    & 0.896    &0    &0    & *  & *  & 0.922    & 0.692   &    0.714& *  & *  \\ \hline
\end{tabular}

\label{table:cov1}
\end{table}
\end{singlespace}
\end{landscape}

\end{document}


\def\spacingset#1{\renewcommand{\baselinestretch}%
{#1}\small\normalsize} \spacingset{1}

\if0\blind
{
  \title{\bf Supplementary material for ``New estimation methods for extremal bivariate return curves"}
  \author{C. J. R. Murphy-Barltrop$^{1*}$, J. L. Wadsworth$^2$ and E. F. Eastoe$^2$\\
\small $^1$STOR-i Centre for Doctoral Training, Lancaster University LA1 4YR, United Kingdom \\
\small $^2$Department of Mathematics and Statistics, Lancaster University LA1 4YF, United Kingdom \\
\small $^*$Correspondence to: c.barltrop@lancaster.ac.uk}
\date{\today}
  \maketitle
} \fi

\if1\blind
{
  \bigskip
  \bigskip
  \bigskip
  \begin{center}
    {\LARGE\bf Supplementary material for ``New estimation methods for extremal bivariate return curves"}
\end{center}
  \medskip
} \fi

\bigskip


\spacingset{1.8}


\section{Angles and lines}
\begin{figure}[tb]
    \centering
    \includegraphics[width=\textwidth]{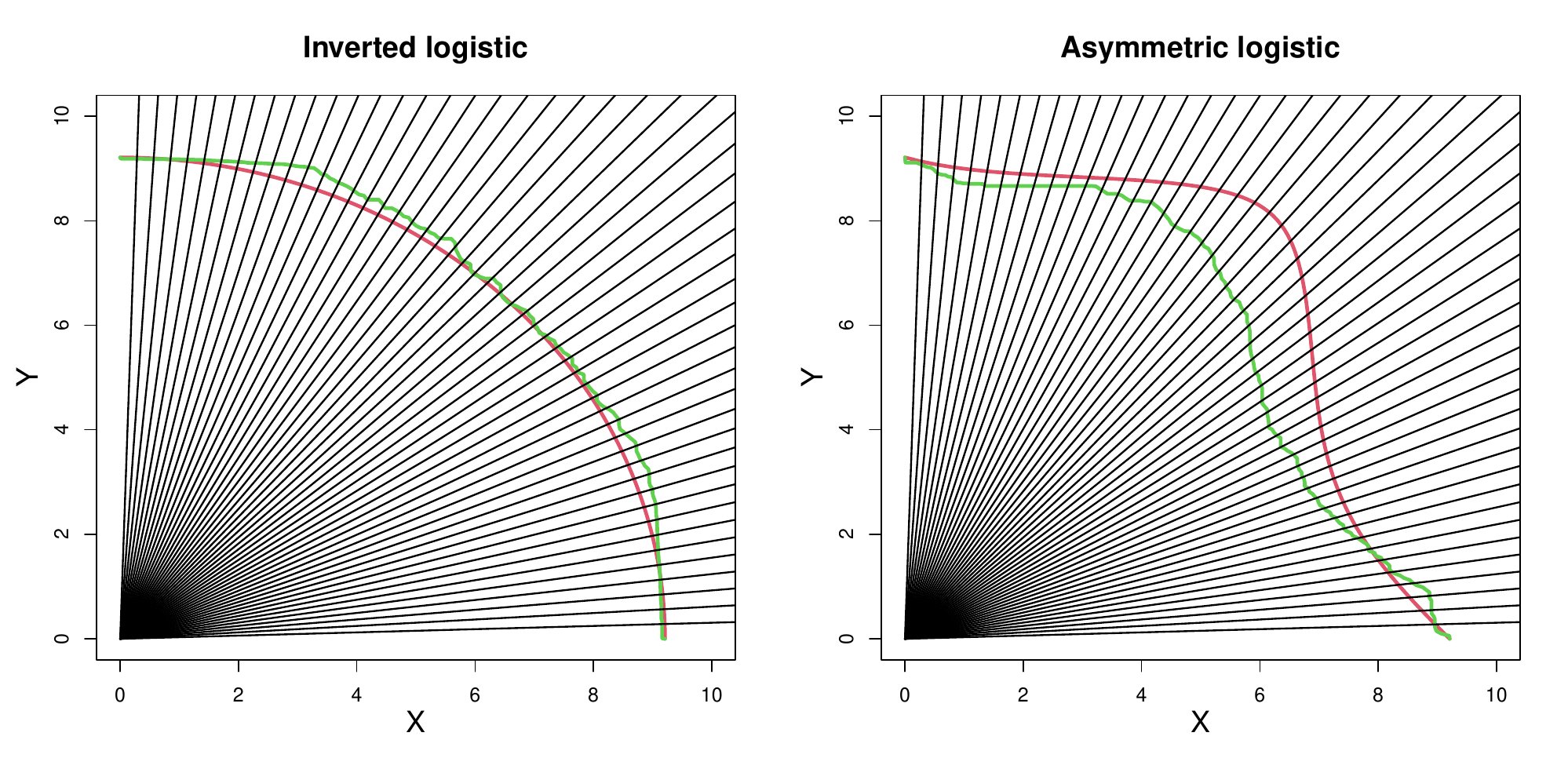}
    \caption{Resulting line segments from a set of angles $\boldsymbol{\Theta}$  with $m=25$ for both copula examples. True and estimated curves given in red and green, respectively.}
    \label{fig:estvstrue2}
\end{figure}

Examples of angles $\theta \in \boldsymbol{\Theta}$ and corresponding line segments $L_{\theta}$ are illustrated in Figure \ref{fig:estvstrue2} for both copula examples discussed in Section 4.1 of the article. From Figure \ref{fig:estvstrue2}, one can observe that each line segment intersects both the estimated and true return curves exactly once; this follows from the definition of a return curve. In this manner, each angle and corresponding line segment represents a common feature of both curves.

\section{Diagnostic example}

\begin{figure}[tb]
    \centering
    \includegraphics[width=\textwidth]{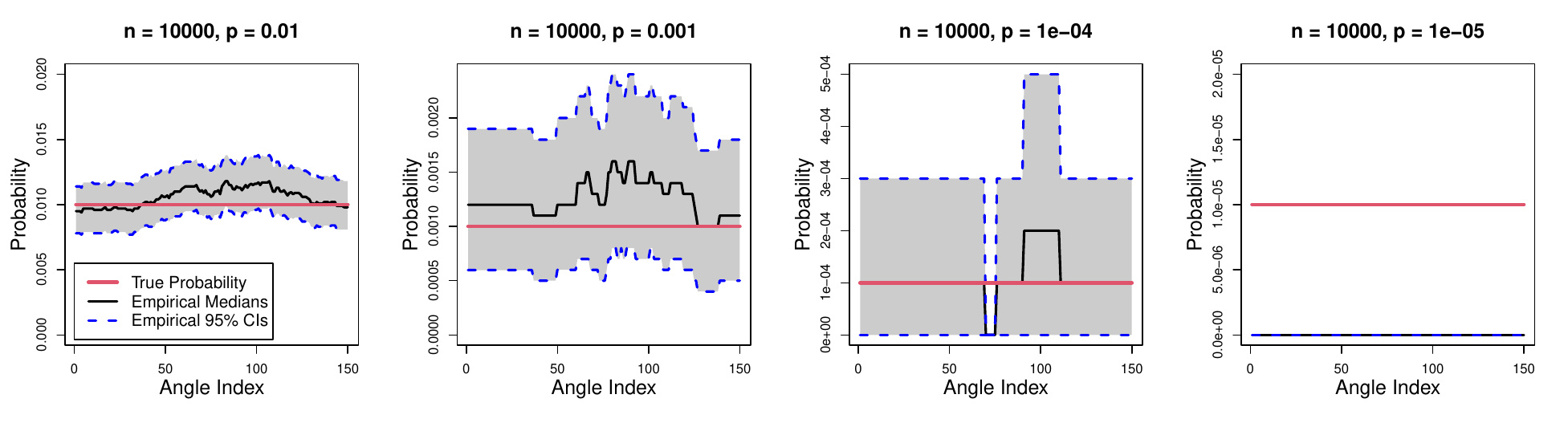}
    \caption{Diagnostic tool for four $p$ values with logistic copula sample and $n=10000$. Red, black, and dotted blue lines represent true values, estimated medians, and estimated 95\% confidence intervals, respectively, for each probability and index $j$.}
    \label{fig:dif_probs}
\end{figure}

In Figure \ref{fig:dif_probs}, we illustrate the diagnostic tool at different probabilities. The sample size is fixed at $n = 10^{4}$ and four probabilities are considered  ($p \in \{10^{-2},10^{-3},10^{-4},10^{-5}\}$).

\section{Summary statistic of integrated absolute difference}

\begin{figure}[tb]
\centering
\includegraphics[width=\textwidth]{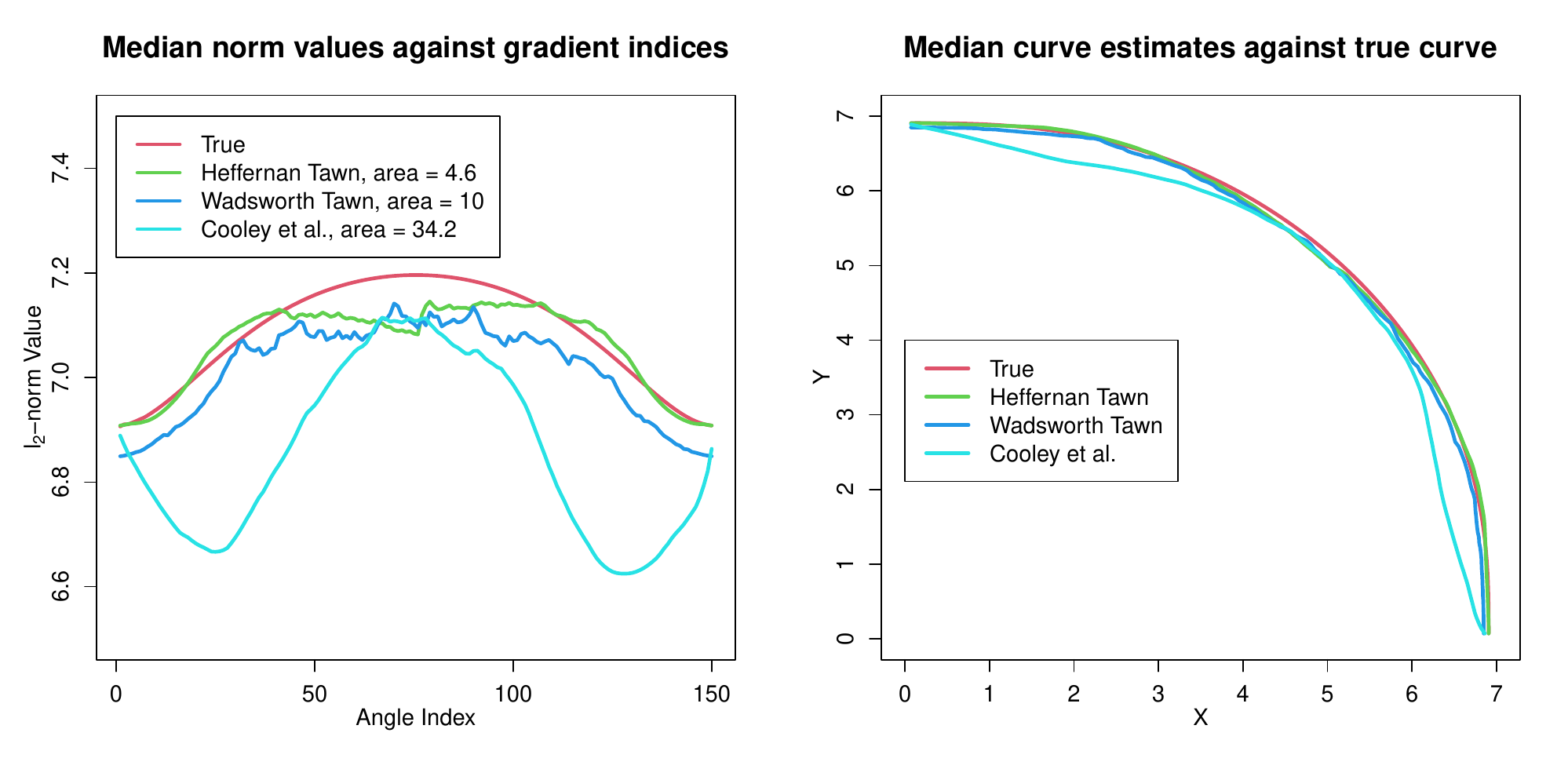}
\caption{Left: median $d_{\theta}$ estimates obtained using $50$ simulated samples from a standard bivariate normal copula with $\rho=0.6$ and true norm values against angle indices. Summary statistics given in plot legend. Right: median curve estimates from each model against true curve for the same example. True values are given in red while the estimated values from the \citet{Heffernan2004}, \citet{Wadsworth2013} and \citet{Cooley2019} models are given in green, dark blue and light blue, respectively.}
\label{fig:summarystat}
\end{figure}

An example of the plot corresponding to the summary statistic used to evaluate bias in return curve estimates is illustrated in the left panel of Figure \ref{fig:summarystat} for a standard bivariate normal copula with $\rho=0.6$, along with the median return curve estimates from each model in the right panel. Letting $d_{\theta_j}$ and $\hat{d}_{\theta_j}$, $j = 1, \ldots, 150$, denote the $l_2$-norm values of true and estimated median curves for angles $\theta_j \in \boldsymbol{\Theta}$, the summary statistic is given by
\begin{equation*}
    A(d,\hat{d}) = \sum_{j=1}^{150} \vert d_{\theta_j}-\hat{d}_{\theta_j} \vert.
\end{equation*} 
The closer this quantity is to zero, the closer the norm values are to the truth and hence the nearer the median curve estimates are to the true curve. Moreover, an area of zero corresponds to an unbiased curve estimate since this implies there is no difference between the estimated and true curves (at the points corresponding to angles in $\boldsymbol{\Theta}$). One can observe the median curve estimate obtained using the \citet{Cooley2019} framework appears to perform poorly for this particular example.

\section{Illustration of procedure for estimating coverage}

\begin{figure}[tb]
\centering
\includegraphics[width=\textwidth]{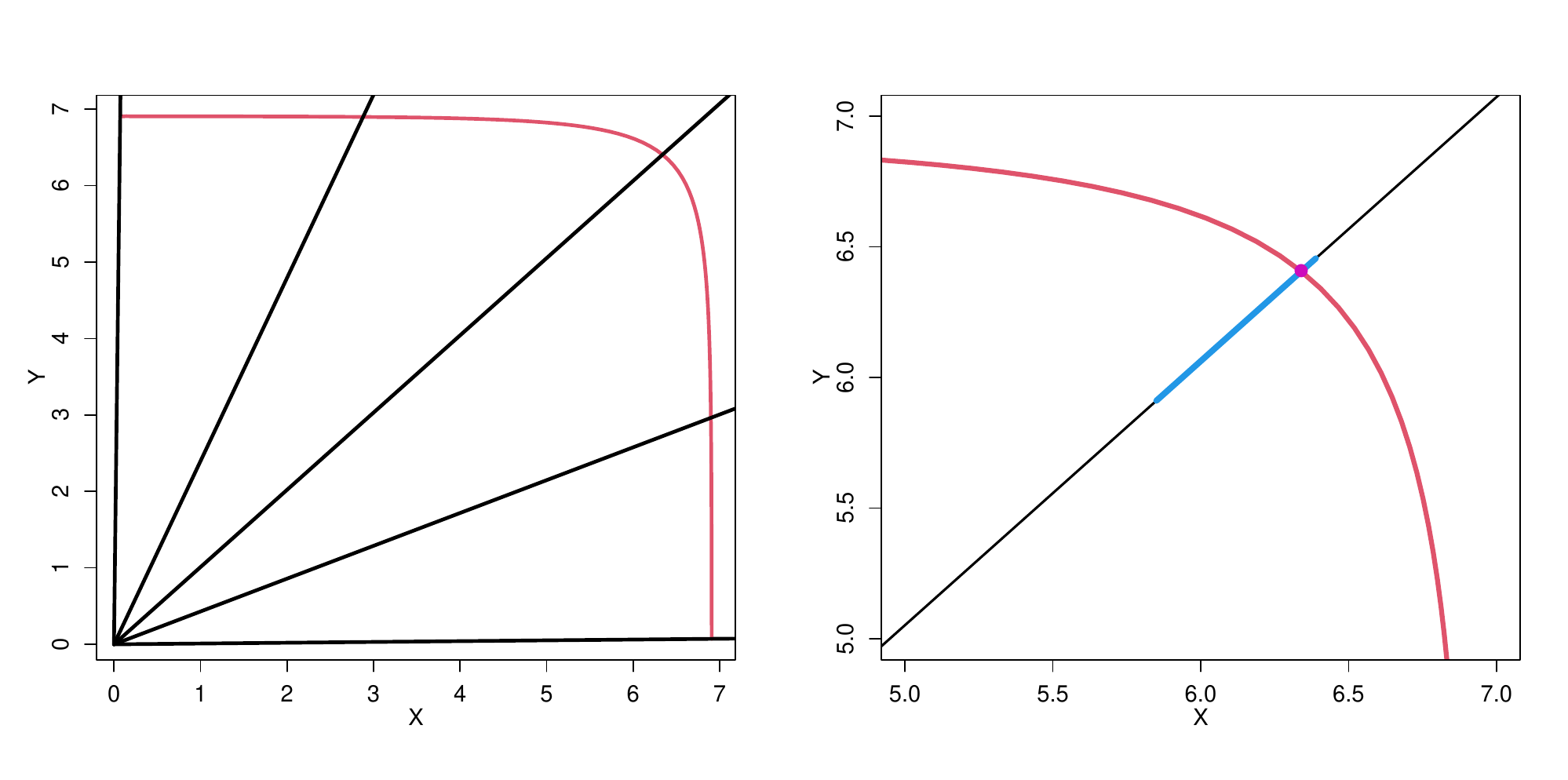}
\caption{Left: Line segments, $L_{\theta}$, corresponding to the angles considered for evaluating coverage. Right: Confidence region computed for one sample at the third angle. The blue line represents the estimated $95\%$ confidence region for $l_2$-norm values along the corresponding line segment. True return curve (red) in both plots obtained from the logistic copula with dependence parameter $0.5$ and $p=10^{-3}$.}
\label{fig:gradient_confidence_region}
\end{figure}

The left panel Figure \ref{fig:gradient_confidence_region} illustrates the five angles used to evaluate coverage. We label these angles 1-5, clockwise from the $y$-axis. Angle 3 is close in value to $\pi/4$, resulting in a line similar in appearance to $y=x$. Angles 4 and 5 are obtained by applying the function $f(\theta) = \pi/2 - \theta$ to angles 1 and 2, respectively, and hence they can, in a sense, be considered symmetric. 

For each simulated sample and angle $\theta$, a confidence region is obtained for estimated coordinates on the line segment $L_{\theta}$. One can record whether the true point at this angle lies within the estimated region; for an unbiased curve estimate, this would be expected $100(1-\alpha)\%$ of the time. An example confidence region is illustrated in the right panel of Figure \ref{fig:gradient_confidence_region}; we note that the true point (pink) lies within the region. Repeating this procedure over the $500$ simulated samples, the proportion of times the true points lie within the estimated confidence regions can be computed, giving an estimated measure of coverage at each of the angles 1-5.

\section{Additional coverage results}
Table \ref{table:cov2} gives the additional coverage results for each copula considered in Section 5 of the article, with $p = 10^{-4}$.



\section{Illustration of reference points and corresponding line segments}
\begin{figure}[tb]
    \centering
    \includegraphics[width=\linewidth]{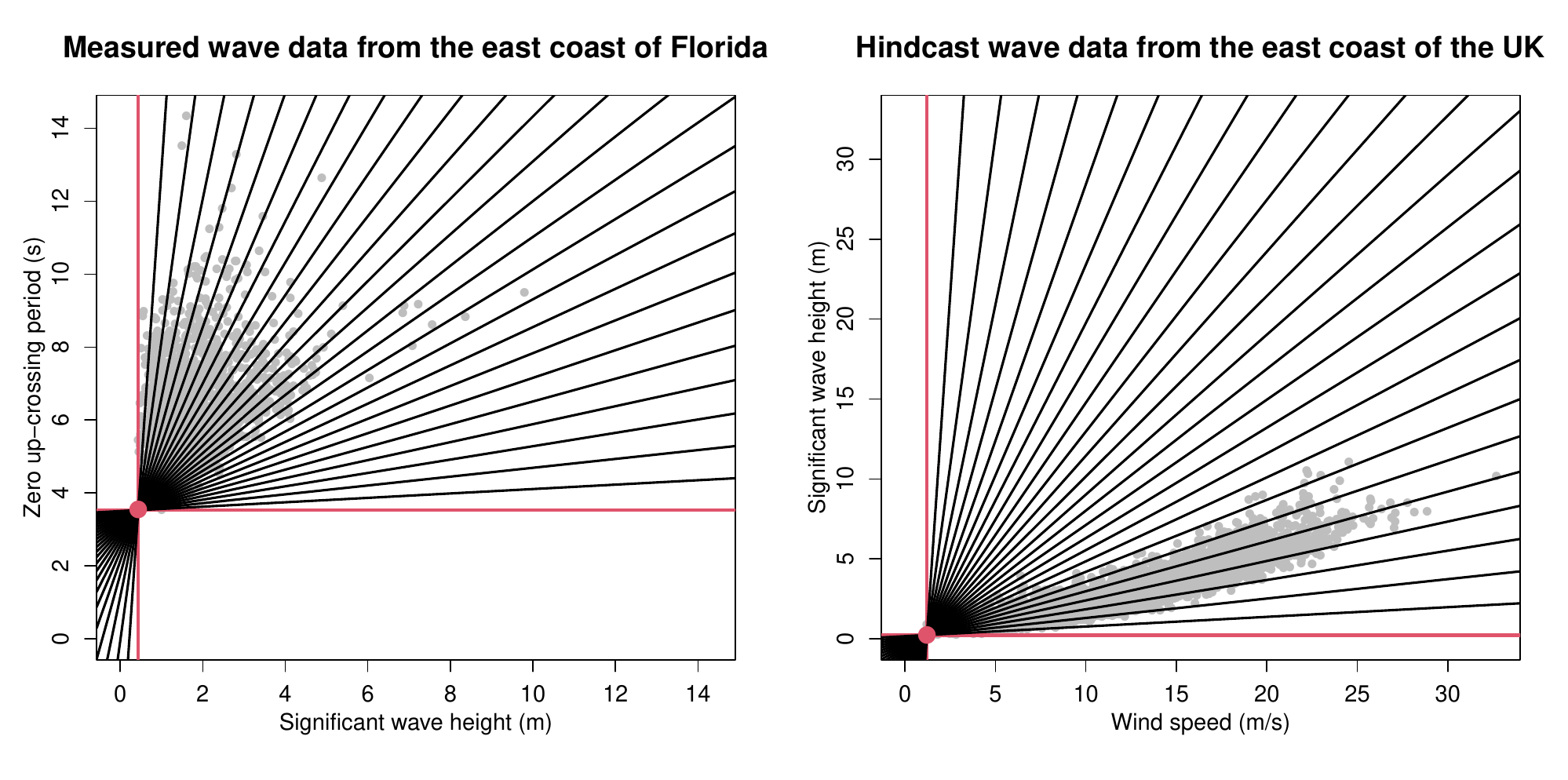}
    \caption{Reference points $(x_0,y_0)$ (red) for measured (left) and hindcast (right) data sets, alongside corresponding line segments (black) $L'_{\theta}$ for $\theta \in \boldsymbol{\Theta}$, with $m = 25$.}
    \label{fig:ref_points}
\end{figure}

Figure \ref{fig:ref_points} illustrations the reference points $(x_0,y_0)$ and corresponding line segments for both data sets considered in the case study of the article. The points at which the line segments intersect return curve estimates are used to quantify uncertainty and define joint survival regions for the diagnostic tool. One can observe that in both cases, defining line segments from the origin $(0,0)$ would result in procedures whereby return curve estimates are evaluated outside of the region where data has been observed: such an evaluation is not meaningful in practice.  

\bibliography{supp.bib}

\begin{landscape}
\begin{singlespace}

\begin{table}[H]
\centering
\caption{Coverage values of $95\%$ confidence regions for $p=10^{-4}$. `HT' and `WT' correspond to the \citet{Heffernan2004} and \citet{Wadsworth2013} models, respectively.}
\begin{tabular}{|c|c|c|c|c|c|c|c|c|c|c|c|}
\hline
\multirow{2}{*}{Copula}                           & Probability & \multicolumn{10}{c|}{$p = 10^{-4}$}               \\ \cline{2-12} 
                                                  & Model       & \multicolumn{5}{c|}{HT} & \multicolumn{5}{c|}{WT} \\ \hline
\multirow{2}{*}{BEV Logistic}                     & Angle    & 1   & 2  & 3  & 4  & 5  & 1   & 2  & 3  & 4  & 5  \\ \cline{2-12} 
                                                  & Coverage    & 0.010     & 0.970    & 0.688   & *  & *  &0.844      & 0.554    & 0.846    & *  & *  \\ \hline
\multirow{2}{*}{BEV Asymmetric Logistic}          & Angle    & 1   & 2  & 3  & 4  & 5  & 1   & 2  & 3  & 4  & 5  \\ \cline{2-12} 
                                                  & Coverage    & 0.592     & 0.936    &0.496    &0.846    &0.954    &0.902& 0& 0.008& 0.718 & 0.948\\ \hline
\multirow{2}{*}{Bivariate Normal 1}               & Angle    & 1   & 2  & 3  & 4  & 5  & 1   & 2  & 3  & 4  & 5  \\ \cline{2-12} 
                                                  & Coverage    & 0.474     & 0.946   & 0.964   & *  & *  & 0.868
& 0.936
& 0.906 & *  & *  \\ \hline
\multirow{2}{*}{Bivariate Normal 2}               & Angle    & 1   & 2  & 3  & 4  & 5  & 1   & 2  & 3  & 4  & 5  \\ \cline{2-12} 
                                                  & Coverage    & 0.942    & 0.928    & 0.936   & *  & *  & 0.936
& 0.938
& 0.934  & *  & *  \\ \hline
\multirow{2}{*}{Inverted BEV Logistic}            & Angle    & 1   & 2  & 3  & 4  & 5  & 1   & 2  & 3  & 4  & 5  \\ \cline{2-12} 
                                                  & Coverage    & 0.850    & 0.928   & 0.944   & *  & *  & 0.860
& 0.946
& 0.946  & *  & *  \\ \hline
\multirow{2}{*}{Inverted BEV Asymmetric Logistic} & Angle    & 1   & 2  & 3  & 4  & 5  & 1   & 2  & 3  & 4  & 5  \\ \cline{2-12} 
                                                  & Coverage    & 0.938     &0.928    &0.952    &0.954    & 0.902   &0.936
& 0.936
& 0.954
& 0.946
& 0.916    \\ \hline
\multirow{2}{*}{Bivariate T 1}                    & Angle    & 1   & 2  & 3  & 4  & 5  & 1   & 2  & 3  & 4  & 5  \\ \cline{2-12} 
                                                  & Coverage    & 0.481     & 0.933   & 0.419   & *  & *  &0.920
& 0.672
& 0.834& *  & *  \\ \hline
\multirow{2}{*}{Bivariate T 2}                    & Angle    & 1   & 2  & 3  & 4  & 5  & 1   & 2  & 3  & 4  & 5  \\ \cline{2-12} 
                                                  & Coverage    & 0.765    & 0.756    & 0.854   & *  & *  & 0.938
& 0.554
& 0.176 & *  & *  \\ \hline
\multirow{2}{*}{Frank}                            & Angle    & 1   & 2  & 3  & 4  & 5  & 1   & 2  & 3  & 4  & 5  \\ \cline{2-12} 
                                                  & Coverage    & 0.902    &0.816    &0    & *  & *  & 0.920
& 0.024
& 0.01 & *  & *  \\ \hline
\end{tabular}
\label{table:cov2}
\end{table}
\end{singlespace}
\end{landscape}